\documentclass[usegraphicx,a4paper]{mn2e}

\usepackage{amsmath,amssymb,amsfonts}
\usepackage{graphicx,color,latexsym} 
\usepackage{natbib}
\bibpunct[;]{(}{)}{,}{a}{}{,} 

\begin{document}

\title{Galaxy cluster's rotation}
\author[]
{
  \parbox[t]{\textwidth}
  {
    M. Manolopoulou${^{1,2}}$\thanks{E-mail: manolo@roe.ac.uk}, M. Plionis${^{2,3}}$
  }
  \vspace*{6pt}\\ 
  \parbox[t]{15 cm}
  {
    $1$ Institute for Astronomy, The University of Edinburgh, Royal Observatory,  
    Blackford Hill, Edinburgh EH9 3HJ, UK.\\
    $2$ Section of Astrophysics, Astronomy and Mechanics, Department of Physics,
    Aristotle University of Thessaloniki, 54 124, Thessaloniki, Greece.\\
    $3$ Instituto Nacional de Astrof\'{\i}sica Optica y Electr\'onica, AP 51
    y 216, 72000, Puebla, M\'exico.
  }
}
\date{\today}
\maketitle

\title{Galaxy cluster's rotation} 
\begin{abstract}
We study the possible rotation of cluster galaxies, developing, testing and applying a novel algorithm which identifies rotation, if such does exist, as well as its rotational centre, its axis orientation, rotational velocity amplitude and, finally, the clockwise or counterclockwise direction of rotation on the plane of the sky.  To validate our algorithms we construct realistic Monte Carlo mock rotating clusters and confirm that our method provides robust indications of rotation. We then apply our methodology on a sample of Abell clusters with $z\lesssim0.1$ with member galaxies selected from the Sloan Digital Sky Survey (SDSS) DR10 spectroscopic data base. After excluding a number of substructured clusters, which could provide erroneous indications of rotation, and taking into account the expected fraction of misidentified coherent substructure velocities for rotation, provided by our Monte-Carlo simulation analysis, we find that $\sim 23\%$ of our clusters are rotating under a set of strict criteria. Loosening the strictness of the criteria, on the expense of introducing spurious rotation indications, we find this fraction increasing to $\sim 28\%$. We correlate our rotation indicators with the cluster dynamical state, provided either by their Bautz-Morgan type or by their X-ray isophotal shape and find for those clusters showing rotation within 1.5 $h^{-1}_{70}$ Mpc that the significance of their rotation is related to the dynamically younger phases of cluster formation but after the initial anisotropic accretion and merging has been completed. Finally, finding rotational modes in galaxy clusters could lead to the necessity of correcting the dynamical cluster mass calculations. 
\end{abstract}


\section{Introduction}
Galaxy clusters are the deepest gravitational wells in the Universe \citep[eg.,][]{well1,well2}, constituting an important ingredient of the Cosmic Web. They form at the interception of filaments and/or walls, where the galaxy density is larger and where infall is strongest \citep[eg.,][]{Rien08,formationreview}. They contain from a few tens to a few thousands of galaxies. The frequency distribution of cluster masses, the mass function, and its evolution have been recognized as a very important cosmological probe that can constrain the current cosmological model \citep[eg.,][]{Borgani08,masscosmology1,masscosmology2}. Consequently, the  accurate calculation of cluster masses is of uttermost importance and a powerful means to cosmological studies. 

Many different approaches have been used to calculate cluster masses; taking advantage of clusters acting as gravitational lenses \citep[eg.,][]{Kneib08,lensmass1,lensmass2,lensmass3}, using the Sunyaev-Zeldovich effect \citep[eg.][]{sz,Birk08,szmass2,szmass1}, assuming hydrostatic equilibrium and calculating the intracluster medium (ICM) temperature from their X-ray emission \citep[eg.,][]{Sarazin08,hydromass2,hydromass1} or using the clusters' galaxy member velocities and assuming dynamical equilibrium \citep[eg.,][]{dynmass2,dynmass1,dynmass3} or not   \citep[eg.,][]{diaf2,diaf1}.  However, not all methods give the same results\citep[eg.,][]{Hoekstra07,Peng09,Donahue14};  each one of them has its own advantages and disadvantages \citep[eg.,][]{massmethods}. Comparing the different methods is important in order to understand the systematic effects that enter in each one and, thus, for estimating more accurate masses \citep[eg.][]{calibr1,calibr2,calibr3}. 

When calculating the cluster mass using the velocities of the individual cluster members, we assume that the cluster is in virial equilibrium; the gravitational potential equals two times the sum of the kinetic energy of the members and that galaxy orbits are roughly isotropic. This does not take into account the possible contribution to the galaxy velocities of a rotational component. Clusters could be rotating due to an initial angular momentum that survives since their formation or due to recent mergers or interactions with close neighbours. Not taking into account the rotation could result in an erroneous dynamical cluster mass, which could ultimately affect the cosmological constraints provided by the cluster mass function. The difficulty to distinguish a rotating cluster from two closely interacting or merging ones is probably the cause for the few early attempts to investigate the rotation of galaxy clusters \citep[eg.,][]{mat,Tov02}. \citet{hwanglee} used the galaxy member velocities to search for indications of rotation and found $\sim$10\% of their cluster sample to be rotating and in dynamical equilibrium (not undergoinng a recent merger). The relevant study of \citet{ham} used galaxy velocities, X-ray spectra of intracluster gas, and distortions of the cosmic microwave background (CMB). \citet{chlub} and \citet{coo} studied the effect of the cluster rotation on the temperature and polarization of the CMB; while other groups have attempted to model the rotation of the intracluster medium \citep{fang,bian}. A particular case study is cluster A2107 which has been found to rotate in multiple studies \citep{mat,oerg,kal}. Recently, a new attempt to study cluster rotation using the SDSS spectroscopic sample concludes that some clusters are indeed rotating \citep{tovm}.

This work aims in identifying the rotation of members of clusters by using their velocities taken from the SDSS DR10 spectroscopic data base \citep{sdss}. We construct a novel algorithm that can identify both the cluster rotation axis and amplitude. We apply the algorithm to selected Abell clusters and seek for correlations between their rotation properties and their dynamical state. When required we use a flat $\Lambda$ cold dark matter cosmology with $H_0=70 h_{70}$ km s$^{-1}$ Mpc$^{-1}$.

The paper is organized as follows: in section 2 we present our rotation identification algorithm and compare it with that of \citet{hwanglee}. In section 3 we test the efficiency of our algorithm, while in section 4 we present our cluster sample, systematic biases and the application of our algorithm. In section 5 we present and discuss our results and we derive our conclusions in Section 6.

\section{Rotation identification}
First, it is important to clarify what we intend in our   work as a rotating cluster. A rotational mode in clusters can be caused by a variety of  mechanisms, among which the anisotropic infall of  material, an initial angular moment of the proto-cluster that  survives virialization, an off-axis merging, etc. Most probably all  of them are related to the initial or secondary bulding of the  cluster and therefore rotation should be expected in many phases during  the cluster formation  process. 

The question however posed in our work is what is the fraction of virialized (or close to) clusters which retain a rotational mode. Therefore, as detailed in the following sections, we have made all  efforts to exclude from our sample clearly interacting clusters, clusters with  multipule components, cluster with detectable substructures in  velocity and projected space. However, clusters in post-merging phase  which are at the process of virialization, but not yet completely virialized,  with no significant substructure indications cannot be easily  distinguished, but in any case we believe that they should probably  be counted among the rotating clusters. For those that  disagree, we also make an effort, through targeted Monte Carlo  simulations, to estimate the fraction of such false detections.

\subsection{Our Method}
\label{ourmethod}
We introduce a method to identify the possible coherent rotation of galaxies in galaxy clusters. The method provides both the rotational velocity amplitude and the orientation of the projected rotation axis, as well as a quantification of the rotation being a true feature or an artefact.

\begin{figure}
 \centering
 \resizebox{\hsize}{!}{\includegraphics[width=0.45\textwidth]{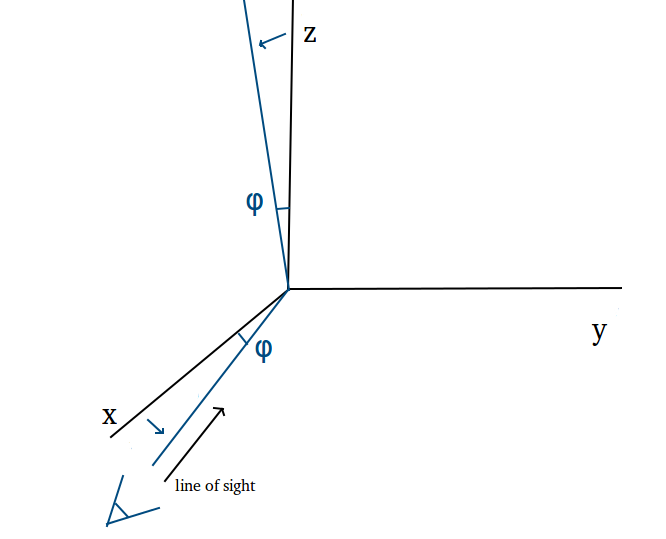}}
 \caption{The triaxial coordinate system and the line of sight    direction (blue line). The $y$-axis remains intact.}
 \label{los}
\end{figure}

\begin{figure}
 \centering
 \resizebox{\hsize}{!}{\includegraphics[width=.5\textwidth]{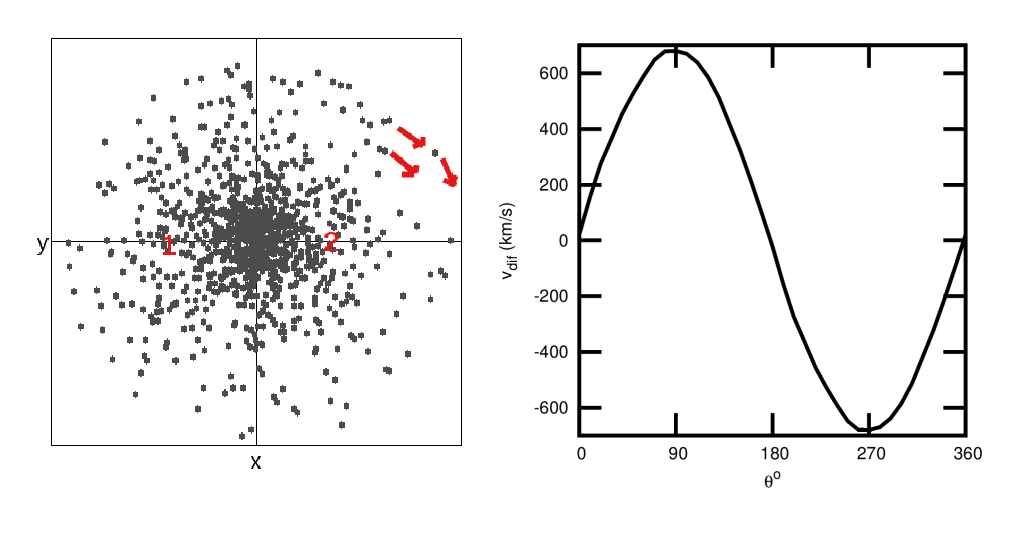}}
 \caption{An illustration of our method. We show a
     Monte Carlo cluster which has been set to counter-rotate with an amplitude $v_{rot}=600$ km/s and with its projection       rotation axis at an angle      $\theta_{rot}=90^{\circ}$ with respect to the North. Our rotation    identification method entails rotating consecutively the galaxies    of the cluster by an angle $\theta$ in the clock-wise direction (as    indicated by the red arrows) and estimating the velocity difference    between the East-West semicircles (details are presented in    the main text). The right-hand panel shows the resulting rotation      diagram, ie., the velocity difference between the two      semicircles against the angle $\theta$.}
 \label{il}
\end{figure}

In order to explain our procedure let us first assume a counter-rotating cluster with constant rotational velocity of 600 km/s, ie., each galaxy member has the same velocity irrespective of its cluster-centric distance. In order to have a realistically ``observed'' cluster we then assign to each galaxy the line-of-sight component of its rotational velocity with respect to the cluster centre (which in our example we set it to be stationary). Starting from the components $v_x,v_y,v_z$ of the  velocity of each galaxy and placing the $y$-axis on the plane of the sky, we calculate its line-of-sight velocity from the relation: $$v_{los}=v_x \cos\phi + v_z \cos(90^{\circ}-\phi)\;,$$  where $\phi$ is the vertical angle between the line of sight and axis $x$ (see Fig. \ref{los}), ie., the $z$-axis is the axis of rotation. For $\phi=0$, the line of sight coincides with the $x$-axis and the cluster rotation axis is perpendicular to the line-of-sight, the ideal case for observing rotation; as the angle $\phi$ increases, we also take into account the $z$-component of the velocity in the line-of-sight velocity. For $\phi=90^{\circ}$, the line of sight coincides with the $z$-axis (see section 3.3 for the effect of $\phi$ on the detection of rotation).
 A visual illustration of our example cluster, which has $\phi=0$ and a horizontal projected rotation axis on the plane of the sky ($\theta_{rot}=90^{\circ}$),  and of our procedure is  provided in Fig. \ref{il}, as detailed below.

The basic idea is to divide the projected distribution of galaxy cluster members  in two semicircles (1 and 2; as shown in the left-hand panel of Fig. \ref{il}), measure the difference of the mean galaxy velocities between the two semicircles, $v_{dif}=\langle v_1\rangle -\langle v_2 \rangle$, and  rotate  consecutively (on the plane of the sky) the galaxy positions by an angle $\theta$ in the clockwise direction (as shown in the left-hand panel of Fig. \ref{il} by the redarrows), repeating the measurement of $v_{dif}$ for each rotation. Consequently, we obtain the velocity difference $v_{dif}(\theta)$ as a function of the angle $\theta$. We will use the graph of $v_{dif}(\theta)$ (right-hand panel of Fig. \ref{il}), which we call rotation diagram, as our primary indication for the presence or not of a rotation mode. 

We need now to relate the observed $v_{dif}(\theta)$ to the true rotation velocity of the cluster. Even in the ideal case of $\phi=0^{\circ}$ we will not observe the whole $v_{rot}$ of each galaxy but as already discussed only its projected component along the line of sight.  We will observe (for those galaxies in the semicircle moving towards the observer) blueshifted velocities with magnitudes which depend on their 3D position in the cluster. For example, if they are located at an angle $\mu$ from the line of sight passing through the centre of the cluster (ie., the angle between the line of sight passing through the centre of the cluster and the cluster radius connecting the centre of the cluster to the galaxy), the observed rotational velocity magnitude of each galaxy will be $v_{obs} =  v_{rot}  \times  \cos(90-\mu)$, where $\mu$ takes values from 0$^o$ to 180$^o$.  The mean $v_{obs}$ of all $N$ galaxies in this projected semicircle will not add up to $v_{rot}$, but to:  $$\langle v_1\rangle \simeq v_{rot}  \times   \sum_{i=1}^N \cos(90^{\circ}-\mu_i)/N$$ Similarly, in the other semicircle it will add up to $\langle v_2\rangle = -\langle v_1\rangle$.  Thus,  $$v_{dif}= \langle v_1\rangle - \langle v_2\rangle =  2 v_{rot}\times \sum_{i=1}^N \cos(90^{\circ}-\mu_i)/N \;,$$ where for convenience we have assumed the same number of galaxies within each semicircle (projected hemisphere) and at symmetric positions to each other (ie., not respecting a realistic volume fill). In such a model configuration we find that: $$\sum \cos(90-\mu_i)/N \sim 0.636 $$ The realistic observational situation, where most of the galaxies are at small angles $\mu$ from the cluster centre due to the larger volume projected, can be estimated directly from our Monte Carlo cluster of Fig. \ref{il}, where we find  $v_{obs} \sim 0.503 v_{rot}$ and therefore $v_{rot} \simeq v_{dif}$. As a result, the rotational velocity of the cluster will be read from the rotation diagram as: $$v_{rot}={\rm MAX}[v_{dif}(\theta)]\;.$$

To be more detailed, our rotation detection procedure  entails rotating on the plane of the sky the galaxy-member positions by an angle $\theta$ starting from the vertical axis clockwise, in the range $0^{\circ}-360^{\circ}$ and with a step, say, of $10^{\circ}$. In our example,  for $\theta=0$, we will not observe any significant velocity difference between the East-West hemispheres;  ideally, at the absence of noise we should obtain $v_{dif}=0$. As $\theta$ increases, the velocity difference should increase until it reaches its maximum value at $\theta=90^{\circ}$. In this case, the galaxies in one semicircle would seem to move away and in the other semicircle would seem to approach us, with respect to the cluster centre. Then as $\theta$ increases to $180^{\circ}$ the amplitude of the rotation signal will decrease  and increase again towards $\theta=270^{\circ}$ until it approaches again $v_{dif}=0$ at $\theta=360^{\circ}$. This behaviour is depicted in the right-hand panel of Fig. \ref{il} which shows the periodic rotation diagram for an ideally rotating cluster with a constant velocity of 600 km/s.

A few interesting and important issues, that will be addressed in the following sections, are as follows.
\begin{itemize}
\item The orientation of the rotational axis with respect to the line   of sight can hamper the detection of a rotational mode, if   such exists (see section 3.1).
\item Based on whether the troughs or the peaks appear first in the rotation diagram, we infer the rotating or counter-rotating nature of the cluster (as an example, in Fig. \ref{il} the cluster is counter-rotating). In an initial irrotational Universe the expectation of course is for a statistically equivalent number of both type of rotating clusters.
\item If a non-rotating cluster has one or more small-sized subgroups with a significant velocity difference with respect to the rest of the cluster then, in the rotation diagram, we may observe narrow peaks or troughs at some angle $\theta$ but not a clearly sinusoidal signal.  However, most such cases will be identified and excluded from our analysis at an early stage (see Sections 4.1.1.and 4.1.2). However, there are cases where global rotation and infalling substructures cannot be easily distinguished: (a) a subgroup occupying a relatively large fraction of the cluster projected area; (b) two significant subgroups of galaxies moving at opposite directions within the cluster potential, although such a case requires fine tunning and thus should be rare. 
\end{itemize}

In general, the expectation for a non-rotating cluster, with no significant infalling substructures, is to have a random rotation diagram (no systematic dependence of $v_{dif}(\theta)$ on $\theta$) with relatively small values of $v_{dif}(\theta)$.

\begin{figure*}
 \centering
 \resizebox{\hsize}{!}{\includegraphics{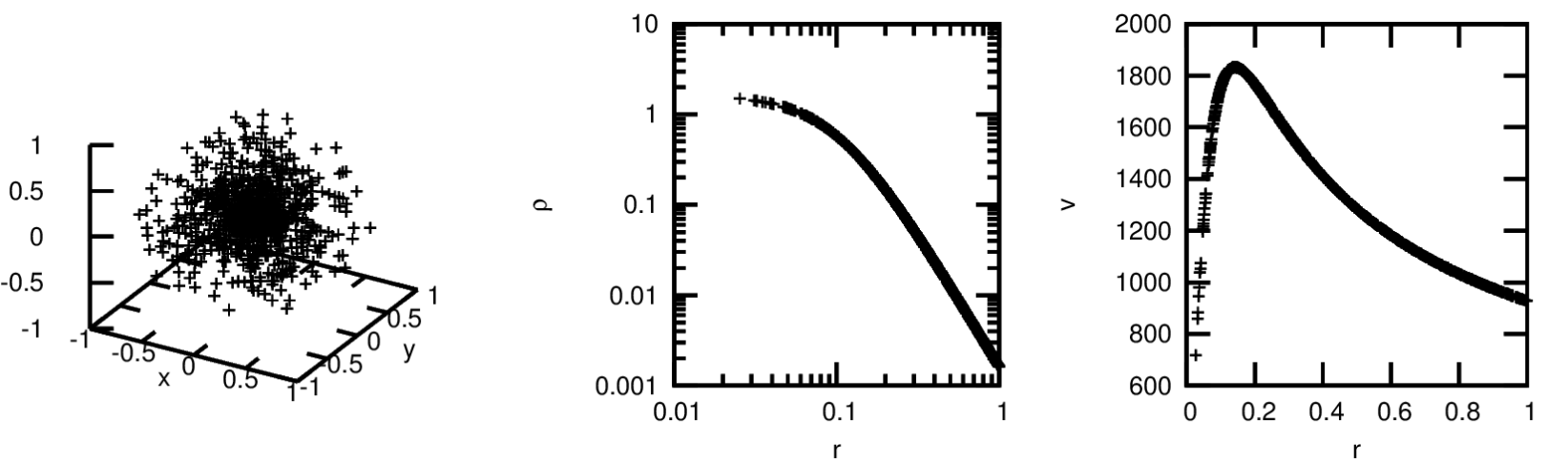}}
 \caption{(a) A Monte Carlo cluster in 3D, (b) the density $\rho$ as a    function of the cluster-centric distance $r$, and (c) the amplitude of    the virial velocity as a function of the distance $r$ from the    cluster centre.}
 \label{eg2}
\end{figure*}

\subsection{The Hwang and Lee method}
Another method to identify cluster rotation, with which we will compare our own, has been proposed by \citet{hwanglee}. They use a sinusoidal relation to compute the angle of the rotation axis, $\Theta_o$, and the rotational velocity $v_{rot}$: 
\begin{equation}
 v_p(v_{rot},\Theta)=v_{sys}+v_{rot}\cdot \sin(\Theta-\Theta_o) \;,
\label{HL}
\end{equation}
where $v_p$ is the predicted radial velocity of each   galaxy due to the cluster rotation, $v_{sys}$ is the peculiar velocity of the cluster and $\Theta$ is projected on the plane of the sky position angle of each galaxy, setting off from North to East. Since in our case we use velocity differences with respect to the cluster mean recessional velocity, we set $v_{sys}=0$.

A $\chi^2$ minimization procedure can be used to determine the best-fitting values of $\Theta_o$ and $v_{rot}$, assuming that the sinusoidal model of equation \eqref{HL} represents well the velocity data. Namely, we use a grid of $\Theta_o$ and $v_{rot}$ values and calculate $\chi^2$ for each pair of parameters: $$\chi^2(v_{rot}, \Theta_o)=\sum_i\frac{(v_{p_i}-v_{los, i})^2}{\sigma_i^2}\;,$$ where $v_{los, i}$ is the observed line-of-sight velocity of every galaxy and $\sigma_i$ its measurement error. 

\section{Validation of our method}
Before applying our method to real galaxy cluster data, we should  validate and confirm that it can provide unambiguous indications of  rotation for the case of realistic clusters and that it can correctly  provide the amplitude of rotation and its axis orientation. To this end, we construct, using the Monte Carlo simulation  method, a virialized cluster with a mass of $4 \times 10^{14}  M_{\odot}$, radius $R_{cl}=1$ Mpc, core radius $r_c=0.1$ Mpc and  having a King's profile density  distribution:
\begin{equation}
\rho(r)=\frac{\rho_0}{\left(1+(r/r_c)^2\,\right)^{3/2}}\;,
\end{equation}
 where $\rho(r)$ is the density included within radius $r$ and  $\rho_0$ is the density in the centre of the cluster. To estimate the  value of $\rho_0$ we use the cluster mass $M_{cl}$,
\begin{equation}
 M_{cl}=\frac{4}{3}
 \frac{\pi R^3 \rho_0}{\left(1+(r/r_c)^2\,\right)^{3/2}}
\label{Mass}
\end{equation}
from which by using $M_{cl}=M(<R_{cl})$ and $r=R_{cl}$ we estimate $\rho_0=6.56 \times 10^{-12}$ kg/km$^3$. Although it is known that the NFW \citep{nfw} profile is a more accurate representation of the dark matter and galaxy density profiles in clusters of galaxies, while the King's profile is applicable mostly to the intracluster gas \citep{king}, it is acceptable to use the latter for the purpose of just testing our methodology. A realization of one such Monte Carlo cluster can be seen in Fig. \ref{eg2}.

Assuming that the cluster is dynamically relaxed (virialized) we can estimate, using the virial theorem, the amplitude of the expected 3D velocity, $v_k$, of each galaxy, which depends on its distance from the cluster centre according to: $v_k^2=G M(r)/2r$, and from equation \eqref{Mass} we obtain: 
\begin{equation}
v_k(r)=\sqrt{\frac{2}{3}\frac{G\pi\rho_0 r^2}{\big(1+(r/r_c)^2\big)^{3/2}}} \;,
\label{above}
\end{equation}
where $M(r)$ is the mass within a sphere of radius $r$. 

Note that each Cartesian component $v_{k_x},v_{k_y},v_{k_z}$ of the virial velocity $v_k(r)$ is assumed to be randomly orientated, while the rotational velocity will have a coherent orientation perpendicular to some rotation axis (in most cases we will assume it to be lying on the plane of the sky). We will further set a counterclockwise direction on the velocity components
   $v_{rot_{x}},v_{rot_{y}},v_{rot_{z}}$, by:
\begin{gather}
 \vec{v}_{rot}\cdot\vec{r}=0 \nonumber \\
 v_{rot}^2=v_{rot_{y}}^2+v_{rot_{x}}^2 \;, \nonumber
\end{gather}
with the first relation ensuring that the coordinate vector $\vec{r}$ is perpendicular to the rotation velocity vector $\vec{v}_{rot}$. The second implies that the $z$-component of the velocity is set 0, in order the rotational velocities to be perpendicular to the rotation axis $z$.

We can now assign to each galaxy a 3D velocity which could be either of:
\begin{itemize}
\item[(a)] a constant rotational velocity (independent of the   cluster-centric distance of each ``galaxy'') having an amplitude, say   a fraction of the maximum virial expectation and a coherent   orientation around a chosen axis, 
\item[(b)] a rotational velocity having as amplitude a constant   fraction of the virial expectation, ie., different at the different   cluster-centric distances and a coherent orientation around a chosen axis,  
\item[(c)] the vectorial sum of the virial expectation and any of the above   two rotational velocity models. This case corresponds to a more   realistic cluster velocity profile and we model it by assigning to  each ``galaxy'' the randomly oriented virial velocity that corresponds to its  distance from the cluster centre, adding vectorially the   rotation velocity.
\end{itemize}

To investigate the systematics related to the realistic observational situation we will attempt to identify the cluster rotation on the plane of the sky. To this end, we project the 3D cluster on one plane, estimating the line-of-sight component of the total (rotational or rotational+virial) velocity of each mock galaxy and imposing its rotation axis to be perpendicular to the line of sight ($\phi=0$, which is the ideal case).  We then apply both algorithms (ours and that of Hwang and Lee) to investigate their performance for both rotational velocity models and for a variety of axis orientations on the plane of the sky. Furthermore, to study sampling effects we simulate mostly two cases;  a cluster with 1000 and a cluster with 50 ``galaxies''.

\subsection{Model (a): constant rotational velocity}
\begin{figure}
 \centering
 \resizebox{\hsize}{!}{\includegraphics{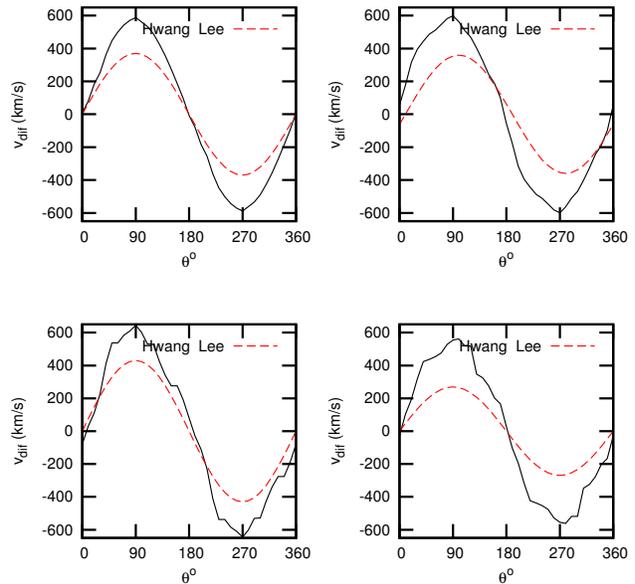}}
 \caption{Comparison of the rotation diagrams of our method    (black continuous line) and of that of Hwang and Lee (red dashed    line) for the cases of $n_{mem}=1000$ (upper panels) and    $n_{mem}=50$ (lower panels).  In the left-hand panels we    present the case of a purely rotational velocity and in the right-hand    panels the case of a total velocity based on the vectorial sum    of the virial expectation and the rotational velocity.     The input rotational velocity has a constant value of    $v_{rot}=540$ km/s.}
 \label{f3}
\end{figure}

Using as input rotational velocity a constant one with  $v_{rot}=540$ km/s (30\% of the maximum virial velocity),  we obtain the results shown in Fig. \ref{f3}, where in the upper panels we present results based on a cluster with $n_{mem}=1000$ and in the lower panels a cluster with $n_{mem}=50$, while in the left-hand panels we present the case of a purely rotational velocity and in the right-hand panels the case of a total velocity based on the vectorial addition of the virial expectation and the rotational velocity. 

For this rotational velocity model we can actually clearly address the issue of how well does each method recover the input rotational velocity (and axis orientation). In Table \ref{fract} we present the output $v_{rot}$ and $\theta_{rot}$ for all four cases shown in Fig. \ref{f3}. For the case of dense sampling, which provides an estimate of the intrinsic performance of the two methods, we find a significant underestimation (by $\sim 35\%$) of the amplitude of $v_{rot}$ by the Hwang and Lee method, and a small ($\lesssim 10\%$) overestimation of $v_{rot}$ by our method. The rotation axis orientation is well recovered by both methods.

\begin{table}
\begin{center}
\caption{Output rotation parameters for our and Hwang and Lee   methods for a Monte Carlo cluster with input parameters:   $v_{rot}=540$ km/s and $\theta_{rot}=90^{\circ}$, analysed in Fig. \ref{f3}.}
\label{fract}
\tabcolsep 6pt
\begin{tabular}{cccccc} 
& & \multicolumn{2}{c}{Our method} & \multicolumn{2}{c}{Hwang \& Lee} \\ \hline
$n_{mem}$ & Rot.model & $v_{rot}$ & $\theta_{rot}$ &$v_{rot}$ &
  $\theta_{rot}$ \\ \hline
1000 & only rot. & 589 & 90 & 370 & 90 \\
1000 & rot+virial& 599 & 90 & 360 & 100 \\
50   & only rot. &645 & 90 & 430 & 90 \\
50   & rot+virial& 562 & 100 & 270 & 90 \\ \hline
\end{tabular}
\end{center}
\end{table}

\begin{figure}
 \centering
 \resizebox{\hsize}{!}{\includegraphics[width=1\textwidth]{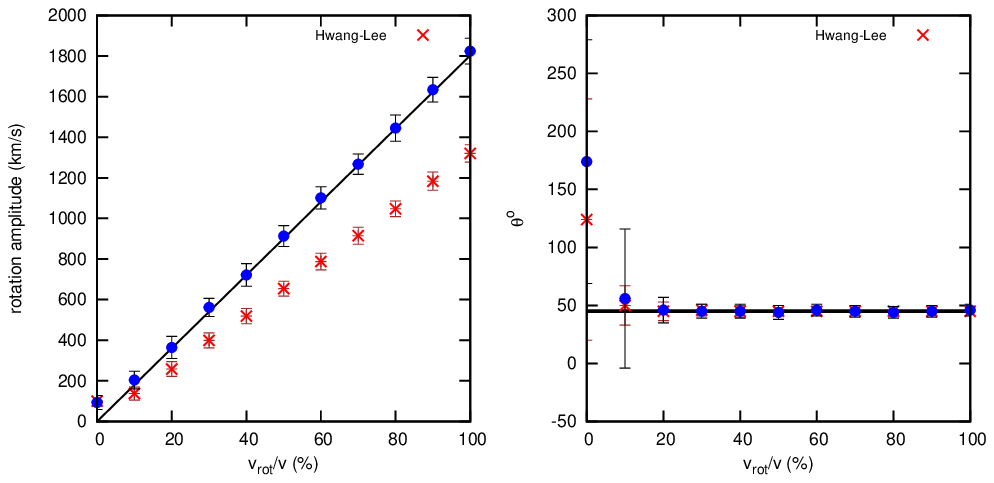}}
\vfill
 \resizebox{\hsize}{!}{\includegraphics[width=1\textwidth]{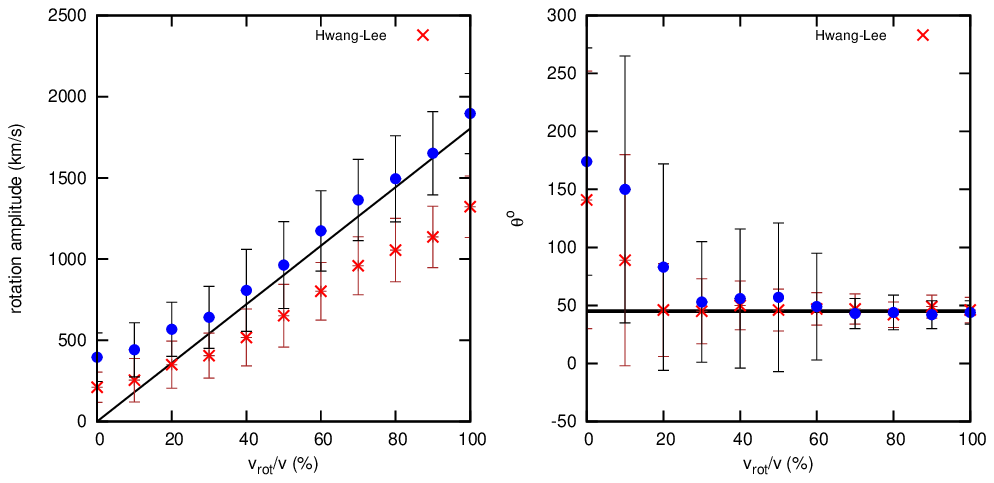}}
 \caption{Recovery of cluster rotational properties as a function of    $v_{rot}/v_{virial}$: left-hand panels: rotation      amplitude, right-hand panels: orientation of the rotation    axis. The black line indicates the input rotation amplitude and    orientation, while the blue and red symbols represent results of    our method and \citet{hwanglee} method, respectively. Upper      panels: for $n_{mem}=1000$ and lower panels: for $n_{mem}=50$.}
 \label{one}
\end{figure}

When we assume sparse sampling, ie., a cluster membership of 50 galaxies,  which is towards the lower limit of the realistic observational cases, we verify that we can still successfully identify the cluster rotational properties but apparently with larger deviations from the input rotational parameters. To substanciate this claim we perform our next important test which is to investigate the rotation identification as a function of the cluster rotational velocity. To this end we simulate sets of 50 Monte Carlo clusters each, all with the same statistical properties, but of which the constant rotational velocity is an increasing fraction of the maximum virial one (from $0\%$ to $100\%$), keeping the same rotation axis orientation ($\theta_{rot}=45^{\circ}$). In order to investigate the convolution of systematics related to the rotation amplitude and to sampling effects, we repeat the procedure for $n_{mem}=1000$ and 50. For each set we calculate the mean and standard deviation of the recovered rotation amplitude and of the orientation of the rotation axis. Their recovery success provides us with the range of cluster parameters for which our method can successfully identify rotation.

In Fig. \ref{one} we present the mean and standard deviation of the recovered rotation amplitudes (left-hand panels) and of the orientation of the rotation axis (right-hand panels) as a function of the ratio $v_{rot}/{\rm MAX}[v_{virial}]$, where   ${\rm MAX}[v_{virial}]=1800$ km/s (see right-hand panel of Fig. \ref{eg2}).  In the ideal case of very good sampling (upper panels), we see that our method correctly recovers the rotation amplitude with negligible uncertainty, except for the case of no rotational velocity where both methods will tend to detect an artificial rotational velocity of $\lesssim 80$ km/s. The already identified problem of the \citet{hwanglee} method, that of underestimating the rotation amplitude, is shown here as well to be true for all $v_{rot}$ being an increasing function $v_{rot}$.  For the sparse sampling cases we have similar overall behaviour as in the dense-sampling cases for both methods but as expected a larger scatter of the resulting rotational parameter values.  In addition we have a larger systematic ovserestimation of $v_{rot}$ by our method, specially for $v_{rot}/v_{virial}\lesssim 0.2$. In general, the uncertainties in the orientation of the rotation axis are quite large for the sparse sampling case, while the Hwang and Lee method performs slightly better in recovering, on average, the correct angle of the orientation axis. 

\subsection{Model (b): fractional rotational velocity of the virial one}
\begin{figure}
 \centering
 \resizebox{\hsize}{!}{\includegraphics{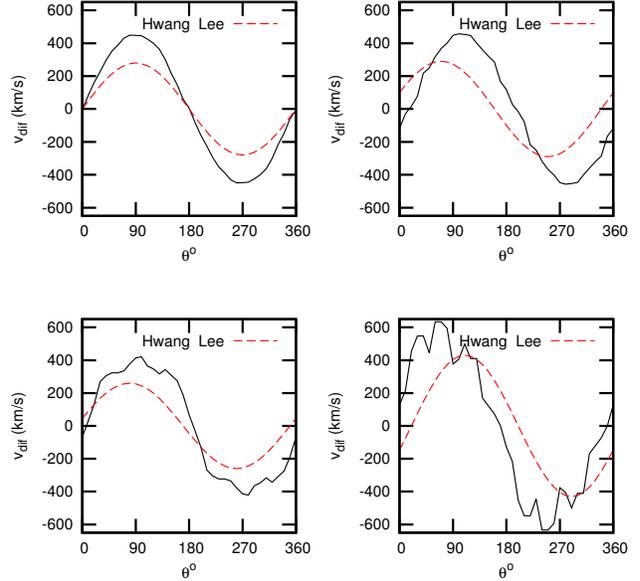}}
 \caption{Comparison of the rotation diagrams of our method    (black continuous line) and of that of Hwang and Lee (red dashed    line) for a rotation model in which $v_{rot}(r)$ is a constant    fraction of the virial velocity at the different cluster-centric    distance and for the cases of $n_{mem}=1000$ (upper  panels) and    $n_{mem}=50$ (lower panels).  In the left-hand panels we    present the case of a purely rotational velocity and in the right-hand    panels the case of a total velocity based on the vectorial sum    of the virial expectation and the rotational velocity.}
 \label{f4}
\end{figure}

For this case we assume a rotation velocity amplitude being a constant percentage (30\%) of the virial; thus $v_{rot}(r)$ depends on the different cluster-centric distances; for example  $v_{rot}(r=1 {\rm Mpc}) \simeq 278$ km/s. Note that for such a rotational velocity field, the output $v_{rot}$ that will be provided by both methods presented in section 2, is an integrated value that depends on the galaxy density and velocity profiles.  In Fig. \ref{f4} we present the rotation diagrams for this case and the output rotational parameters for both methods and for both $n_{mem}$ cases are shown in Table 1. Again, we see the same $v_{rot}$ underestimation of the \citet{hwanglee} method, discussed in section 3.1 for the case of a constant rotational velocity field, which implies that such an understimation is independent of the rotation velocity model.

The analysis of the performance of the two methods when $v_{rot}(r)$ is an increasing fraction of $v_{virial}(r)$ has provided qualitatively similar results as those of Fig. \ref{one} and thus we do not present the corresponding figure.

\begin{table}
\begin{center}
\caption{Output rotation parameters for our and Hwang and Lee   methods for a Monte Carlo cluster with input parameters:   $v_{rot}(r)=0.3 v_{virial}(r)$ km/s and $\theta_{rot}=90^{\circ}$,   analysed in Fig. \ref{f4}.}
\label{fract2}
\tabcolsep 6pt
\begin{tabular}{cccccc} 
& & \multicolumn{2}{c}{Our method} & \multicolumn{2}{c}{Hwang \& Lee} \\ \hline
$n_{mem}$ & Rot.model & $v_{rot}$ & $\theta_{rot}$ &$v_{rot}$ &
  $\theta_{rot}$ \\ \hline
1000 & only rot. & 450 & 80 & 280 & 90 \\
1000 & rot+virial& 457 & 100 & 273 & 90 \\
50   & only rot. & 421 & 100 & 256 & 90 \\
50   & rot+virial& 499 & 110 & 430 & 110 \\ \hline
\end{tabular}
\end{center}
\end{table}

\subsection{Effects of rotational axis orientations with respect to the line-of-sight}
We wish to investigate the effect of different orientations of the 3D rotational axis with respect to the line-of-sight on the rotation identification by our method. In order not to mix the outcome of this test with issues related to sampling effects, we simulate a cluster with dense sampling (ie. having 1000 members). We set initially the rotation axis at a perpendicular position with respect to the line of sight and consequetively rotate the rotation axis with respect to the vertical position, so that it forms an angle $\phi$ with the line of sight in the interval ($0^{\circ},90^{\circ}$) until it is aligned with the line of sight.
 
\begin{figure}
 \centering
 \resizebox{\hsize}{!}{\includegraphics[width=1.0\textwidth]{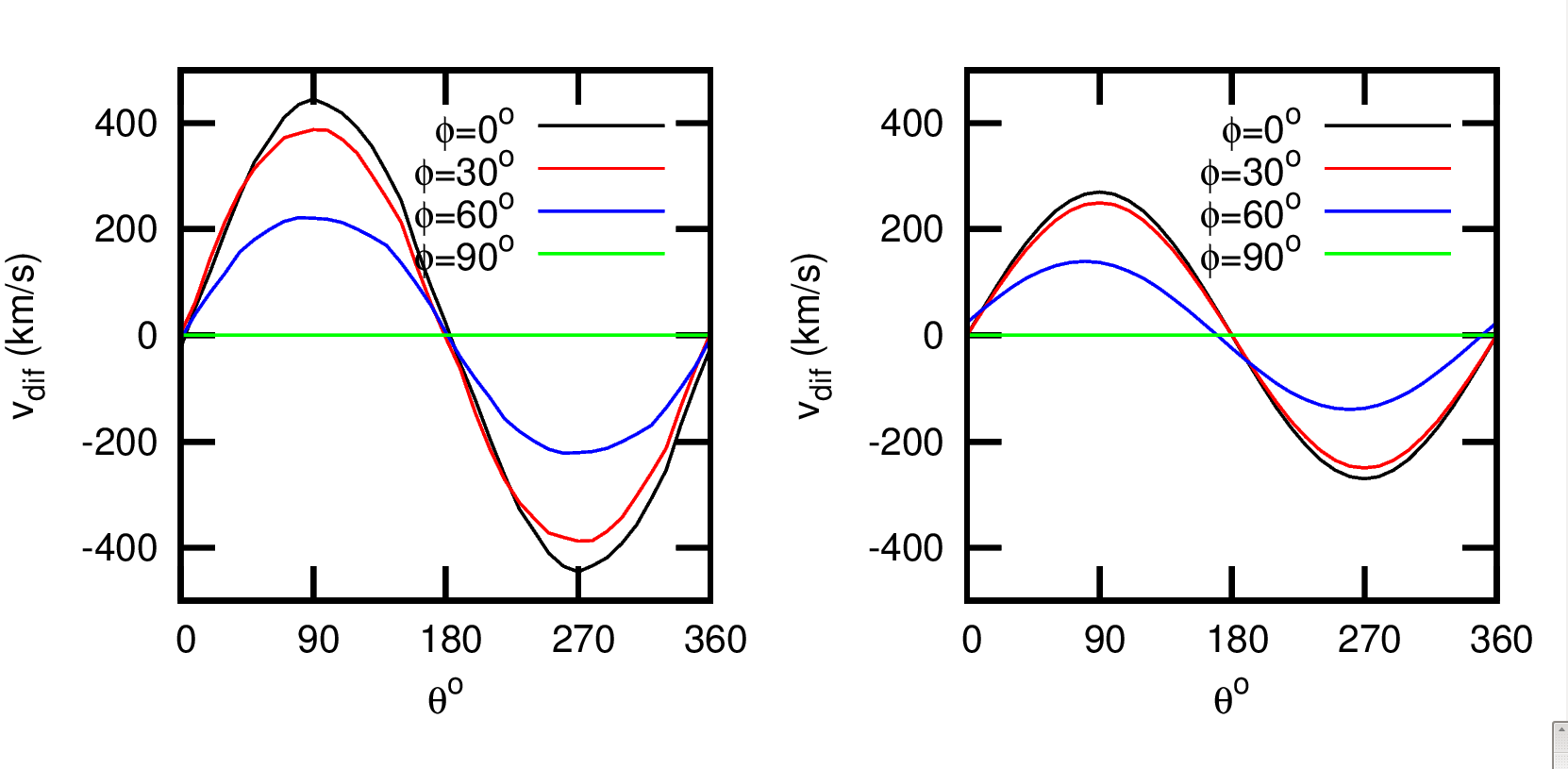}}
 \vfill
 \resizebox{\hsize}{!}{\includegraphics[width=1.0\textwidth]{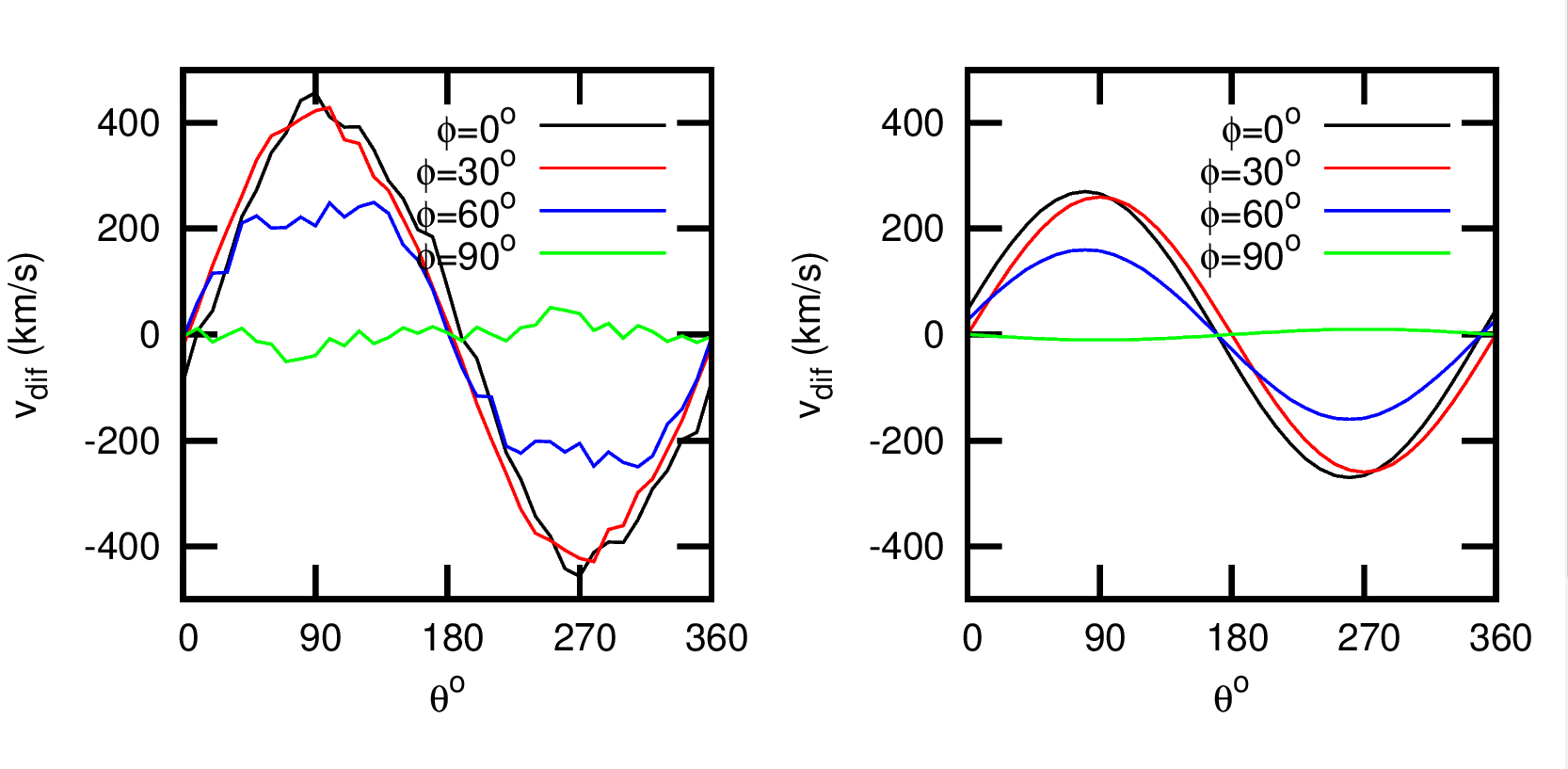}}
 \caption{
The rotation diagram for the cluster of Fig. \ref{eg2} with a rotational velocity 30\% of the virial (which provides an integrated 3D rotational velocity of $\sim 450$km/s), as the    rotation axis shifts from perpendicular to parallel to the line of    sight, ie.,  $\phi\in [0^{\circ}, 90^{\circ}]$. Upper panels    correspond to the    ideal case where only rotational velocities are assigned, while    lower panels correspond to the more realistic case of a 3D    vectorial sum of virial and rotational velocities. Also, the left-hand    panels correspond to the results of our method, while right panels to    results of the \citet{hwanglee} method.}
 \label{phir}
\end{figure}

We apply this procedure using the rotation model b (section 3.2)  and for two cases, an ideal where we assign only the corresponding rotational velocity to each mock galaxy, and a more realistic where we also vectorially add the corresponding randomly orientated virial velocity. The results for different values of the orientation of the rotation axis with respect to the line of sight are shown in Fig. \ref{phir} for both methods, ours and that of \citet{hwanglee}. The upper panels corresponds to the ideal case while the lower panels to the more realistic one.

As expected, the rotation signal becomes weaker (the rotation amplitude decreases) as the angle $\phi$ increases. The counter-rotating direction of rotation is apparent due to the occurrence of the peak at $\theta_{rot}=90^{\circ}$ (ie., because $<180^{\circ}$). At $\phi\sim 90^{\circ}$ the rotation cannot be identified, as the rotation component of the velocity of the galaxies is perpendicular to the line of sight and thus it cannot be observed. Both methods give a flat rotation diagram in   this case, as they should. We also see that the rotation amplitude in the ideal positional case ($\phi=0^{\circ}$) is accurately recovered by our method while it is underestimated by $\sim 35\%$ when using the \citet{hwanglee} method. Their method, as in the ideal 2D case which is presented later, appears to have problems in recovering the correct input rotation amplitude for any value of $\phi$. Furthermore, the accuracy of the recovered rotation axis angle is quite good for both methods (it decreases slightly with the increase of $\phi$). Similar results we recover also in the more realistic case (lower panels), with the addition that the rotation signal becomes practically undetectable for $\phi\gtrsim 60^{\circ}$.

Therefore, we conclude that we will miss a fraction of intrinsically rotating clusters due to axis orientation effects, even in the best case of dense sampling. If we make the reasonable assumption that the rotation axis of each cluster is randomly orientated with respect to the line of sight, the fraction of missed rotating clusters  can be estimated as the ratio of the solid angle that corresponds to an angle $\delta\phi\sim 30^{\circ}$ to the solid angle of the whole sphere, ie., $$f_{missed}\simeq \sin^{2}(\delta\phi/2)\simeq 0.10$$ This should be considered a strict lower limit to the expected number of missed rotating clusters, since sparser sampling will detariorate the detectability of rotation.

\subsection{Conclusions on the method performance}
We can conclude the following from the extended Monte Carlo simulation analysis of the performance of the rotation indentification procedure that: 
\begin{itemize}
\item On the limit of dense-sampling our method recovers very well both the input rotation amplitude and the orientation of the rotation axis, while the \citet{hwanglee} method although accurately identifies the  rotation axis orientation, it systematically understimates the rotation amplitude by $\sim 35\%$, \item On the limit of sparse-sampling our method systematically   overestimates by   $\lesssim 10\%$ the input rotation amplitude. Identifying the   correct orientation of the rotation axis is more   demanding with typical uncertainties being as large as $\sim   50^{\circ}$. The \citet{hwanglee} method performs better than our   method in identifying the correct axis orientation, 
\item One should not expect to recover correctly the   rotation characteristics if the rotation velocity is $\lesssim 10-15\%$   of the virial velocity (ie., typically $\lesssim 200$ km/s) and the   sampling of the   cluster members is low ($<50$ galaxies/cluster), ie.,    the richer the cluster the easier the rotation signal   can be identified and the more accurately the rotation properties   can be recovered, 
\item A fraction of rotating clusters will be missed due to the   orientation of the rotation axis being close to the line of sight. A   crude estimate indicates this fraction to be at least 10\%. 
\end{itemize}

\section{Data Analysis}
\subsection{Cluster and Galaxy Data}
 Our original sample consists of all Abell/ACO clusters  \citep{Abell89} of richness class $R\ge 1$ and distance class 4 or 5  that are located in the SDSS survey area and have more than 50  galaxies with SDSS DR10 spectroscopy within a rest-frame radius of  2.5 $h_{70}^{-1}$ Mpc from  the cluster centre and within a redshift  separation of $\delta z=0.01$ from the central cluster redshift (as  provided by the NASA/IPAC Extragalactic Database). After excluding a few clusters  that are affected by the survey borders we are left with a sample of  103 Abell clusters, presented in Table \ref{ttabl}. 

Note that the  line of sight velocity of each galaxy is given by \citep{danesevlos}: $$v_{los}=c\times  \frac{z_{gal}-z_{cl}}{1+z_{cl}}\;\;,$$  where $z_{cl}$ is the  cluster redshift, while the cluster velocity dispersion  is then provided by: $$\sigma_v= \sqrt{\sum_{i=1}^N \frac{v_{los,i}^2}{N-1}}\;\;,$$ where $N$ is the total number of galaxies used in the estimation.

\begin{table*}
\begin{center}
\tabcolsep 8pt
\caption{The Abell clusters of our sample. From left to right the columns correspond to: Abell   names, redshifts, celestial coordinates, BM type and a measure of the cluster richness, provided by the number of bright ($M>M^{*}$) members within the 1.5$h^{-1}_{70}$ Mpc   radius.}
\label{ttabl}
\begin{tabular}{rcrrcrrcrrcr}\hline
\textbf{Cluster} & \textbf{$z$} & \textbf{RA$(^{\circ})$} &
\textbf{Dec$(^{\circ})$} & \textbf{BM} & \textbf{$N_{*}$} &
\textbf{Cluster} & \textbf{$z$} & \textbf{RA$(^{\circ})$} &
\textbf{Dec$(^{\circ})$} & \textbf{BM} & \textbf{$N_{*}$}\\\hline
85 & 0.0551 & 10.408 & -9.343 & 1 & 3 & 1749 & 0.0573 & 202.385 & 37.626 & 2 & 2 \\
87 & 0.0550 & 10.757 & -9.793 & 3 & 7 & 1767 & 0.0703 & 204.001 & 59.212 & 2 & 3 \\
168 & 0.0450 & 18.791 & 0.248 & 2.5 & 10 & 1773 & 0.0765 & 205.536 & 2.248 & 3 & 5 \\
257 & 0.0703 & 27.247 & 13.982 & 2.5 & 4 & 1775 & 0.0717 & 205.482 & 26.365 & 1 & 8 \\
279 & 0.0797 & 29.093 & 1.061 & 1.5 & 4 & 1780 & 0.0786 & 206.159 & 2.883 & 3 & 7 \\
426 & 0.0179 & 49.652 & 41.515 & 2.5 & 2 & 1795 & 0.0625 & 207.252 & 26.585 & 1 & 3 \\
659 & 0.1005 & 126.020 & 19.404 & - & 6 & 1809 & 0.0791 & 208.329 & 5.154 & 2 & 7 \\
690 & 0.0788 & 129.810 & 28.840 & 1 & 3 & 1827 & 0.0654 & 209.561 & 21.707 & 2 & 2 \\
724 & 0.0933 & 134.575 & 38.573 & 2.5 & 5 & 1831 & 0.0615 & 209.793 & 27.991 & 3 & 3 \\
727 & 0.0951 & 134.780 & 39.422 & 3 & 5 & 1864 & 0.0870 & 212.076 & 5.447 & 2 & 1 \\
957 & 0.0360 & 153.489 & 0.915 & 1.5 & 2 & 1904 & 0.0708 & 215.533 & 48.556 & 2.5 & 7 \\
1024 & 0.0734 & 157.073 & 3.761 & 2 & 2 & 1913 & 0.0528 & 216.716 & 16.676 & 3 & 12 \\
1035 & 0.0684 & 158.030 & 40.209 & 2.5 & 9 & 1927 & 0.0948 & 217.759 & 25.663 & 1.5 & 5 \\
1066 & 0.0699 & 159.850 & 5.173 & 2 & 6 & 1939 & 0.0881 & 219.309 & 24.834 & 2.5 & 4 \\
1137 & 0.0349 & 164.404 & 9.6156 & 3 & 4 & 1983 & 0.0436 & 223.183 & 16.746 & 3 & 5 \\
1168 & 0.0906 & 166.859 & 15.913 & 2.5 & 7 & 1986 & 0.1185 & 223.289 & 21.913 & 3 & 3 \\
1169 & 0.0586 & 167.028 & 43.946 & 3 & 8 & 1991 & 0.0587 & 223.626 & 18.631 & 1 & 6 \\
1173 & 0.0759 & 167.297 & 41.579 & 2.5 & 2 & 2022 & 0.0578 & 226.082 & 28.423 & 3 & 6 \\
1185 & 0.0325 & 167.699 & 28.678 & 2 & 7 & 2028 & 0.0777 & 227.388 & 7.527 & 2.5 & 5 \\
1187 & 0.0749 & 167.915 & 39.578 & 3 & 5 & 2029 & 0.0773 & 227.745 & 5.762 & 1 & 3 \\
1190 & 0.0751 & 167.943 & 40.845 & 2 & 4 & 2030 & 0.0919 & 227.850 & -0.073 & 1.5 & 2 \\
1203 & 0.0751 & 168.489 & 40.294 & 2.5 & 5 & 2034 & 0.1130 & 227.555 & 33.528 & 2.5 & 12 \\
1205 & 0.0754 & 168.343 & 2.511 & 2 & 5 & 2040 & 0.0460 & 228.188 & 7.430 & 3 & 9 \\
1213 & 0.0469 & 169.121 & 29.260 & 3 & 7 & 2048 & 0.0972 & 228.825 & 4.382 & 3 & 12 \\
1228 & 0.0352 & 170.374 & 34.326 & 2.5 & 8 & 2061 & 0.0784 & 230.314 & 30.655 & 3 & 7 \\
1235 & 0.1042 & 170.733 & 19.626 & 2 & 5 & 2062 & 0.1122 & 230.400 & 32.067 & 2 & 4 \\
1238 & 0.0733 & 170.742 & 1.092 & 3 & 6 & 2063 & 0.0349 & 230.758 & 8.639 & 2 & 2 \\
1291 & 0.0527 & 173.019 & 56.024 & 3 & 2 & 2065 & 0.0726 & 230.678 & 27.723 & 3 & 8 \\
1307 & 0.0832 & 173.200 & 14.524 & 2 & 6 & 2067 & 0.0739 & 230.812 & 30.906 & 3 & 6 \\
1318 & 0.0578 & 173.993 & 55.033 & 2 & 8 & 2069 & 0.1160 & 230.991 & 29.891 & 2.5 & 8 \\
1345 & 0.1095 & 175.295 & 10.689 & 3 & 2 & 2079 & 0.0690 & 232.020 & 28.878 & 2.5 & 8 \\
1346 & 0.0975 & 175.293 & 5.689 & 2.5 & 10 & 2089 & 0.0731 & 233.172 & 28.016 & 2 & 5 \\
1358 & 0.0809 & 175.694 & 8.223 & 2 & 5 & 2092 & 0.0669 & 233.331 & 31.149 & 2.5 & 2 \\
1367 & 0.0220 & 176.123 & 19.839 & 2.5 & 10 & 2107 & 0.0411 & 234.950 & 21.773 & 1 & 5 \\
1371 & 0.0687 & 176.355 & 15.507 & 3 & 3 & 2122 & 0.0661 & 236.122 & 36.127 & 2.5 & 0 \\
1377 & 0.0514 & 176.741 & 55.739 & 3 & 8 & 2124 & 0.0656 & 236.247 & 36.061 & 1 & 0 \\
1383 & 0.0597 & 177.038 & 54.622 & 3 & 4 & 2142 & 0.0909 & 239.567 & 27.225 & 2 & 4 \\
1385 & 0.0831 & 177.019 & 11.556 & 3 & 0 & 2147 & 0.0350 & 240.572 & 15.895 & 3 & 13 \\
1408 & 0.1102 & 178.443 & 15.388 & 2.5 & 2 & 2151 & 0.0366 & 241.313 & 17.749 & 3 & 18 \\
1424 & 0.0768 & 179.391 & 5.038 & 3 & 6 & 2152 & 0.0410 & 241.343 & 16.449 & 3 & 2 \\
1436 & 0.0658 & 180.117 & 56.255 & 3 & 8 & 2175 & 0.0951 & 245.095 & 29.915 & 2 & 7 \\
1474 & 0.0801 & 181.988 & 14.955 & 3 & 9 & 2197 & 0.0308 & 247.044 & 40.907 & 3 & 12 \\
1516 & 0.0769 & 184.739 & 5.239 & 2.5 & 6 & 2199 & 0.0302 & 247.154 & 39.524 & 1 & 10 \\
1526 & 0.0799 & 185.535 & 13.739 & 3 & 7 & 2244 & 0.0968 & 255.683 & 34.047 & 1.5 & 8 \\
1541 & 0.0893 & 186.861 & 8.840 & 1.5 & 9 & 2245 & 0.0850 & 255.687 & 33.530 & 2 & 12 \\
1552 & 0.0858 & 187.458 & 11.741 & 2 & 3 & 2255 & 0.0806 & 258.129 & 64.093 & 2.5 & 9 \\
1650 & 0.0838 & 194.693 & -1.753 & 1.5 & 5 & 2356 & 0.1161 & 323.938 & 0.123 & 2.5 & 3 \\
1656 & 0.0231 & 194.953 & 27.981 & 2 & 15 & 2399 & 0.0579 & 329.386 & -7.794 & 3 & 8 \\
1658 & 0.0850 & 195.295 & -3.436 & 2.5 & 2 & 2428 & 0.0851 & 334.061 & -9.350 & 2 & 3 \\
1663 & 0.0843 & 195.694 & -2.518 & 2 & 2 & 2644 & 0.0693 & 355.291 & 0.094 & 2 & 3 \\
1668 & 0.0634 & 195.964 & 19.265 & 2 & 3 & 2670 & 0.0762 & 358.543 & -10.405 & 1.5 & 10 \\
1691 & 0.0721 & 197.847 & 39.201 & 2 & 11 &     &        &         &        &       &  
\end{tabular}
\end{center}
\end{table*}

\subsubsection{Clearing projection effects}
Once we have selected our cluster sample and before we apply our rotation algorithm, we wish to clean each cluster of possible galaxy outliers and projection effects. Indeed, projected galaxies along the line of sight, but separated in velocity space would be a source of noise and could hide or erroneously enhance a rotation signal.

To this end, we plot for each cluster the relative to the cluster centre galaxy velocity frequency distribution, which has a mean value of zero. We expect that a virialized cluster should have a roughly Gaussian frequency distribution of line-of-sight velocities \citep{gauss1,gauss2,gauss3}. Therefore a Gaussian is fitted to the data using the usual $\chi^2$ minimization procedure. Then, outliers are identified as those galaxies with velocities $>3\sigma$ away from the mean, which then are not considered in the rotation analysis.

\subsubsection{Separating substructures}
\begin{figure}
 \centering
 \resizebox{\hsize}{!}{\includegraphics[width=0.4\textwidth]{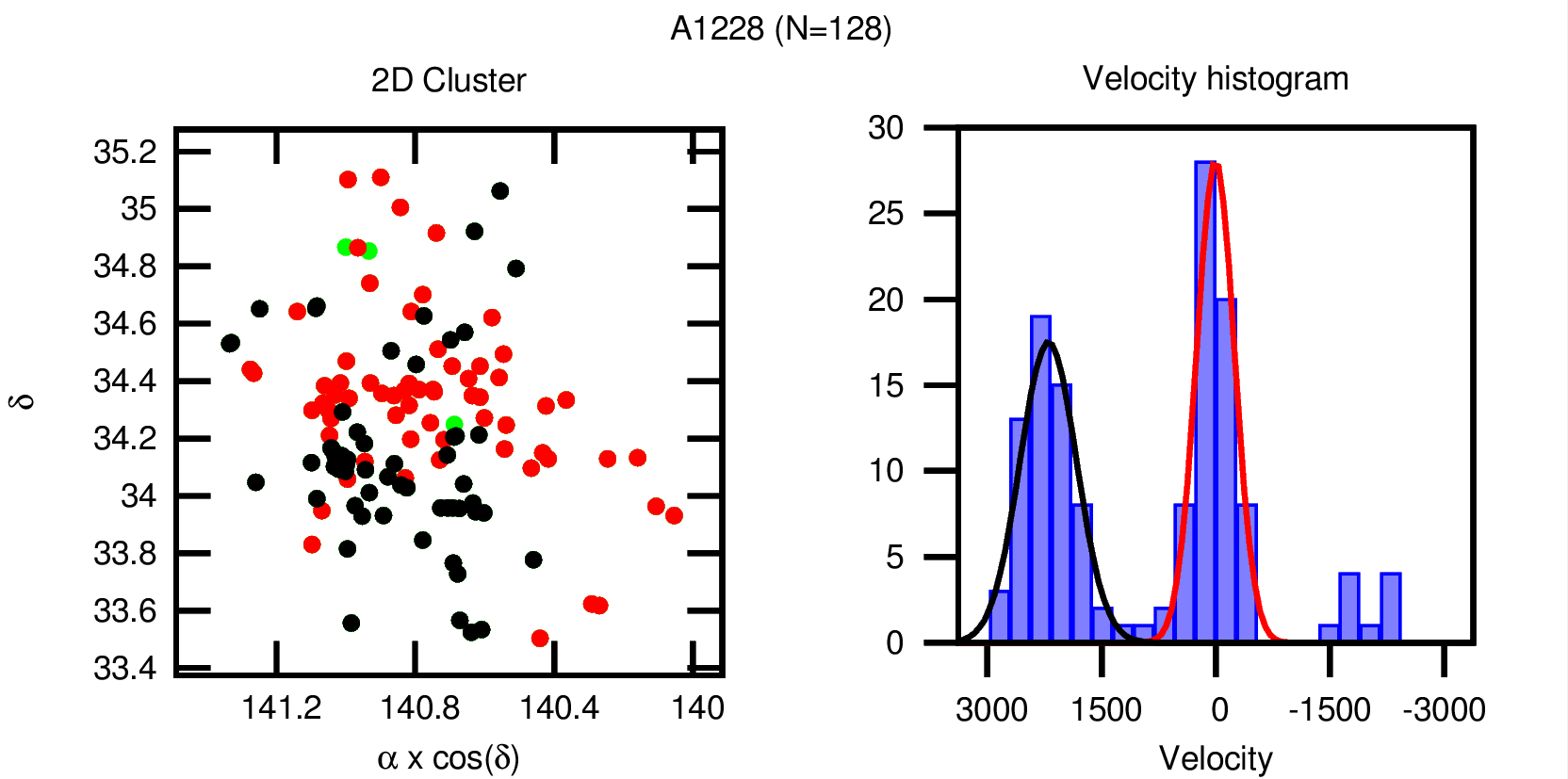}}
 \caption{Left-hand Panel: the projected distribution of galaxies in    the A1228 cluster. Different colours indicate the galaxies    belonging in the three different groups. Right-hand Panel: the    relative velocity distribution of the A1228 galaxies. The colour of    the fitted Gaussian is that of the corresponding members seen in    the left-hand panel. The smallest group is depicted with green in the    left-hand panel.}
 \label{A1228}
\end{figure}

Furthermore, projected groups along the line of sight, but separated in velocity space, or substructures which have coherent infall velocities towards the parent cluster centre, could provide an erroneous rotation signal. In many occasions, it is easy to identify such cases due to either the fact that in projection the substructures are clearly spatially separated from the main cluster, or in other occasions where the different subclusters are clearly separated in velocity space but may appear as a unique cluster in projection. We have carefully inspected all of our clusters and identified those with significant subclumps and each was separately analysed for rotation. As an example, we show in Fig. \ref{A1228} the case of Abell 1228. The left-hand panel shows the projected galaxy distribution,  within a radius of 2.5 $h_{70}^{-1}$ Mpc, which appears as a typical centrally concentrated cluster, while the right-hand panel shows the relative velocity distribution which reveals three clearly separate subclumps (each separated by $\delta v\sim 1900$ km/s from the central one) projected along the line of sight. Had we analysed the whole ``cluster'', without separating the individual subclumps, we would have found a clear and strong signal of rotation. The two larger clumps have more than 50 members each and were separately analysed for rotation (and as we will see they do show strong rotation indications; see the Appendix).

This procedure was finally applied to the following clusters,  A659, A1035, A1228, A1291, A1775, A2067, A2197, A2245, A2255 and A2152, which were found to be composed of two or more subclusters,  increasing our total sample of clusters under study to 110.  However, only in five we managed to perform the separation  procedure effectively (A1035, A1291, A1228, A1775, and A2152), with details being presented in the Appendix\footnote{  We wish to note that had we not separated these clusters they would all have shown significant and strong indications of rotation, exactly due to the coherent velocity differences of the subclusters. Nevertheless, in some cases, as we will see further below, one or even both separated subclusters show true evidence of rotation.}. The rest were tagged as being dominated by substructures and were not included in our final analysis.

Finally, one must also ask what happens if the number of substructure member galaxies is a relatively small fraction of the whole cluster and/or the infall velocity is not as large as to be clearly separated in velocity space. Could such non-rotating clusters be erroneously identified as rotating? In section 4.3.4 we present extensive Monte Carlo simulations tailored to answer such a question and provide the expected fraction of false rotation detections. However, as a first step in excluding such cases, we have investigated in detail all clusters, even if their galaxy velocity distribution appears Gaussian, and we have tagged as being  dominated by substructures those clusters which are spatially dispersed with no clear central core, or clusters for which we found in the literature strong and unambiguous substructure indications (eg., \citet{einasto,krause}). The following clusters fall in this category: A257, A1137, A1187, A1190, A1205, A1346, A1358, A1383, A1385, A1424, A1474, A1749, A1780, A1986, A2028 and A2069.

 \subsubsection{Cluster richness and mass}
In order to have a more accurate determination of the cluster richness with respect to the original Abell's richness class  and to investigate possible richness dependencies of our results, we calculate for each cluster the number of bright galaxies, $N_{*}$, ie. those with $M>M^{\ast}$ in the $r$-band (with $r   \leq 17.7$), using the luminosity function of \citet{lumfun} with the K and evolutionary corrections of \citet{kcor}. In order to check for obvious systematic effects we have tested whether the number of bright galaxies, $N_{*}$, correlates with the cluster redshift. No such correlation was found indicating that $N_{*}$ is a redshift-free indication of the cluster richness and thus of the cluster mass.

Another indicator of the cluster mass is the cluster velocity dispersion, which is related to the mass via the virial theorem. A large velocity dispersion indicates a large cluster mass. Note however that cluster merging and significant cluster substructures can increase the measured velocity dispersion, but in this case it is not necessarily related to the cluster mass but to the highly unrelaxed cluster state.  

One would expect the above two indicators of the cluster mass (velocity dispersion and richness $N_{*}$) to be correlated and indeed they have a Spearman correlation coefficient of $R_s\simeq 0.43$ with a probability of this correlation being random of ${\cal P}\simeq 3\times 10^{-6}$ (velocity dispersion and richness are estimated out to 2.5 $h^{-1}_{70}$ Mpc).

\subsubsection{Cluster dynamical state}
We also wish to investigate whether the possible cluster rotation is related to the cluster dynamical state. If, for example, the anisotropic accretion of matter along large-scale filaments entails infall with non-zero angular momentum, one may expect enhancement of rotational modes towards the cluster centre. To investigate this possibility we will use two well known indicators of the cluster dynamical state; their Bautz-Morgan (BM) type and the shape of their ICM X-ray profile.

The BM type \citep{bm} of Abell clusters is an indication of their morphology and thus of their dynamical state. It can be numerically characterized by a value increasing from one to three (1-3) with two intermediate categories, which we index here as 1.5 and 2.5, respectively. The dynamical youth increases in the same order (or the dynamical evolution inversely), with BM type 1 indicating the most dynamically evolved cluster (spherically symmetric, centrally concentrated and cD dominated) while with BM type 3 the most loose, asymmetric and thus unrelaxed cluster.

Similarly, we will use all the available X-ray cluster images to characterize their dynamical state. We define the X-ray profile parameter, $X_p$, which can take three possible values, $X_p=1$ for roughly spherically symmetric and smooth X-ray emission profiles (virialized and dynamically evolved),  $X_p=2$ for asymmetric and/or distorted profile  (dynamically young) and $X_p=0$ if the X-ray image is not available. The main source of the X-ray images used come from the {\em Einstein} observations \citep{ein}. In total, we have available X-ray images for 49 out of the 110 Abell clusters of our sample.

Since both previously discussed parameters should reflect the cluster dynamical state, they should be correlated. Indeed, we find that the two parameters  correlate nicely and provide a Spearman correlation coefficient of $R_s\simeq 0.53$ and a  probability of this correlation being random of ${\cal P}\simeq 10^{-4}$.

\subsection{Application of our algorithm}
The rotation analysis is performed using galaxies within either a circular region around the cluster centre, having a radius of 1.5$h^{-1}_{70}$ or 2.5$h^{-1}_{70}$ Mpc, or within circular rings of different widths. The latter because we wish to investigate whether the cluster's possible rotation signal comes from the outskirts or the central cluster regions, but also because the central regions are affected more severely by projection effects that could contribute in weakening an existing rotation signal.

By identifying the cluster regions, if any, that show a rotational signal we may get hints as to which is the mechanism producing it. In virialized clusters one may expect that virial relaxation would have erased any initial rotational mode. However, in oblate-like clusters [although clusters appear to be mostly prolate-like \citep[eg.,][]{shape,Basilakos00}]  the collapse along their minor axis may retain and even enhance some initial rotation. On the other hand, if in dynamically young clusters the rotation is caused by interactions and merging, one should expect only the cluster outskirts to show more prominent rotational indications. Excluding from our analysis the cluster central regions, where projections along the line of sight are more severe, may be helpful in this respect. 

Summarizing, we will investigate the cluster rotation in each cluster using four different angular configurations:
\begin{enumerate}
 \item the circular area within 1.5$h^{-1}_{70}$  Mpc radius;
 \item the circular ring within 0.3-1.5$h^{-1}_{70}$ Mpc;
 \item the circular area within 2.5$h^{-1}_{70}$  Mpc radius;
 \item the circular ring within 0.5-2.5$h^{-1}_{70}$ Mpc.
\end{enumerate}

\begin{figure}
 \centering
 \resizebox{\hsize}{!}{\includegraphics[width=0.3\textwidth]{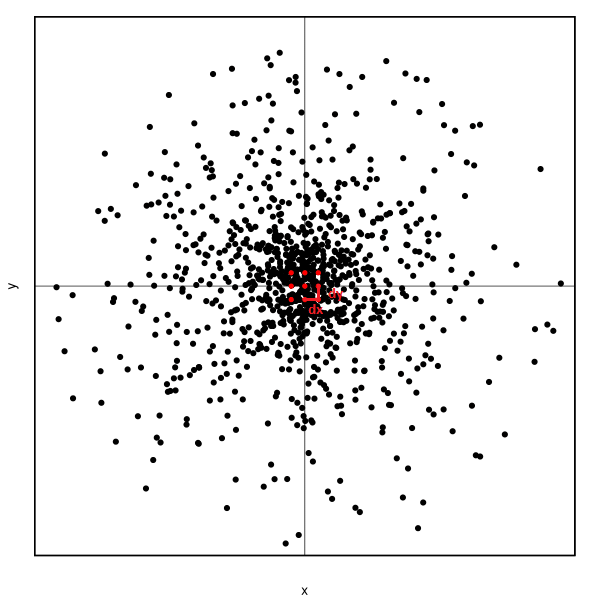}}
 \caption{A mock cluster with the nine different rotation centre candidates shown as red dots.}
 \label{points}
\end{figure}

A further issue that could be important in identifying a rotational mode in clusters is the selection of the true rotational centre, if such exists. We therefore apply our rotation identification procedure using nine different possible centres, forming a rectangle around the nominal centre of the cluster (Fig. \ref{points}). The separation between the consecutive centres is usually 5\% of the cluster radius (in some cases we used 10\%, depending on the size of the cluster). We finally choose that centre as our optimum rotational centre for which the smooth sinusoidal ``ideal rotation'' curve (see section 4.3.1) fits best the data rotation curve, ie. centre which corresponds to the minimum $\chi^2$ value (see Fig. \ref{points} for an example).

\subsection{Identification of significant rotation} In order to make a decision whether a cluster has a significant rotational mode or not we will use the combination of two tests, which
consist of: 
\begin{itemize}
\item[(a)] comparing the distribution of relative velocities, in each   of two hemispheres and for each rotation angle $\theta$, using a   Kolmogorov-Smirnov two-sample test, and 
\item[(b)] comparing the data rotation curve separately with an   ``ideal rotation curve'' and a ``random rotation curve'' suitably   estimated for each cluster, by using the usual $\chi^2$   statistic. If $v(\theta_i)$ and $v_m(\theta_i)$ are respectively the   data and model mean velocity difference between the two semispheres   at a rotation angle $\theta_i$, we define:
\begin{equation}
  \chi^2=\sum_{\theta=0}^{360}\frac{\left(v_\theta-v_{m,\theta}\right)^2}{\sigma_{\theta}^2+
    \sigma_{m,\theta}^2}
  \label{chi3}
\end{equation}
with the data rotation curve uncertainty given, at each rotation angle $\theta$, by: 
\begin{equation}
\sigma_{\theta}^2=\left(
\frac{\sigma_{v,1}}{\sqrt{n_{1}}}\right) ^2+ \left(
\frac{\sigma_{v,2}}{\sqrt{n_{2}}} \right)^2 \;,
\end{equation}
where $\sigma_{v,1}$ and $\sigma_{v,2}$ are the velocity dispersions and $n_{1}$ and $n_{2}$ are the number of galaxies in semicircles 1 and 2, respectively, at each rotation angle $\theta$. The uncertainty of the model rotation curve, $\sigma_{m,\theta}$, is  provided by the scatter among the different Monte Carlo realizations of the random or ideal rotation curves.
\end{itemize}

\subsubsection{Test 1: ideal versus random rotation curves}
In order to build the ideal rotation and random rotation curves we follow the following recipe: For each cluster we identify the angle at which the maximum velocity difference   (MAX$[v_{dif}]\equiv v_{rot}$) is observed in its rotation diagram ($\theta_{rot}$). This angle splits the cluster in two semispheres; and to the galaxies of each we attach a velocity $v_{los}=v_{rot}/2$ and $v_{los}=-v_{rot}/2$, respectively. We then apply our rotation identification algorithm on this new configuration to produce the ``ideal rotation'' diagram with which we compare the data rotation diagram, quantifying the goodness of fit by the $\chi^2$ statistic (equation \ref{chi3}), which we name $\chi^2_{id}$. A value of $\chi^2_{id}/df \lesssim 1$ (where $df$ are the degrees of freedom, in our case the number of steps in $\theta$) shows that the data rotation curve is well represented by the ideal one.

We also construct for each cluster a rotation curve which corresponds to that of a random distribution of velocity residuals. To this end, we shuffle the galaxy line-of-sight velocities randomly while keeping the same galaxy coordinates. Then, our rotation identification algorithm is applied and this process is repeated 10000 times. The final ``random rotation curve'', is the average over all the realizations, while the scatter $\sigma_{v_{r_i}}$ around the mean is also estimated. Finally, we determine the $\chi^2$ statistic between the data and random rotation curves, which we tag $\chi^2_r$. 

We can now select the candidate rotational clusters as those for which $$\chi_{id}^2 \ll \chi_r^2 \;,$$ ie., those for which the ideal rotation curve fits the data rotation curves significantly better. If the opposite occurs then the cluster is likely not rotating. Therefore the ratio $\chi_{id}^2/\chi_r^2$ is a useful parameter for assessing rotation or not. 

\subsubsection{Test 2: KS two sample test}
We also apply the Kolmogorov-Smirnov two-sample test to the distributions of the relative velocities of the galaxies of the two cluster semicircles for each angle $\theta$. This test practically calculates the probability, $P_{KS}$, that the two relative velocity distributions have the same parent distribution. The bigger the probability the more likely it is that the two distributions are mutually consistent. For a rotating cluster we expect a significant difference between the two velocity distribution, and the corresponding $P_{KS}$ probability limit is taken, somehow arbitrarily, to be $P_{KS}=0.01$ (ie., values lower than this limit are taken to signify significantly different distributions).

\subsubsection{Final criteria for a rotating cluster}
We therefore have four criteria that can be used to deduce a significant or not cluster rotation, which we call the strict criteria, and are as follows:
\begin{itemize}
 \item $\chi_{id}^2/df$ between the real and ideal rotation curve,    which should be less or equal to 1 for a rotating cluster, 
 \item $\chi_r^2/df$ between the real rotation curve and random curve,    which should be $> 1$ for a rotating cluster,
 \item $\chi_{id}^2/\chi_r^2$, which should be ideally $\ll 1$ for a    rotating cluster, but practically we take it to be $\le 0.2$,    and \item the Kolmogorov-Smirnov probability, $P_{KS}$, between the    galaxy relative velocity distributions of the two semicircles of    maximum difference, which should be: $P_{KS}<0.01$.
\end{itemize}
These criteria can be relaxed to provide a less secure identification of rotation. For example, we also checked for clusters fullfilling only the following two criteria: $\chi_{id}^2/\chi_r^2<0.4$ and  $P_{KS}<0.01$. We call these the {\em loose} criteria for cluster rotation.

\subsubsection{The fraction of false detections}
We wish to address the issue of what is the fraction of false detections of rotation according to the above selected criteria, when there is no intrinsic rotation present. Two such different possibilities of false detections exist: 
\begin{itemize}
\item due to shot noise, related to small number statistics, and 
\item due to the presence of an unidentified substructure that has a   coherent infall velocity with respect to the cluster mean, and which   can be erroneously assessed as rotation. The substructures that could   still remain unidentified after the procedure discussed in Section   4.1.2 are those which cannot be spatially or dynamically separated   (ie., those which are near the cluster centre and which do not have   a large infall velocity).  
\end{itemize}

To investigate these possibilities we simulate 1000 Monte Carlo clusters, according to the basic recipe of section 3 with either 50 or 100 members, which is the membership range more susceptible torotation misidentification. 

To address the first possibility we assign to the mock galaxies only virial velocities and find only a small fraction of our mock clusters showing a false rotational signal. Under the strict criteria we find a $\sim$1.9\% false detection rate for clusters with either $N=50$ or $100$ members. Using the loose criteria, the corresponding fraction is $\sim$ 4.4\%. These fractions are small enough to allow us to conclude that shot-noise effects are unimportant for the size of clusters considered in this work.

In order to address the second possibility, we introduce a subclump which contains between 10\% and 28\% of the main cluster members, positionally placed on one of the projected quadrants of the cluster at a distance of 420 $h^{-1}_{70}$ kpc from the cluster centre and having as a mean infall velocity a fraction (50\% or 100\%) of the cluster virial velocity dispersion (note that we randomly assign to each substructure member an infall velocity having the above mean and a standard deviation of 500 km/sec). The range of these parameters where selected after a number of trials in order to mimic cases where the $3\sigma$ clipping of the member velocity distribution or the clear positional identification of the substructure would have failed to identify the substructure as such. In Fig. 10 we present the results as the probability of misidentifying an infalling substructure (with the previously discussed characteristics) for cluster rotation as a function of substructure richness (in percentage of main cluster membership). We see that for the case where such substructures exist the probability of them being misidentified as a cluster rotation is between $\sim 0.05 -0.3$ depending on the substructure richness and infall velocity. 

\begin{figure}
 \centering
 \resizebox{\hsize}{!}{\includegraphics[width=.5\textwidth]{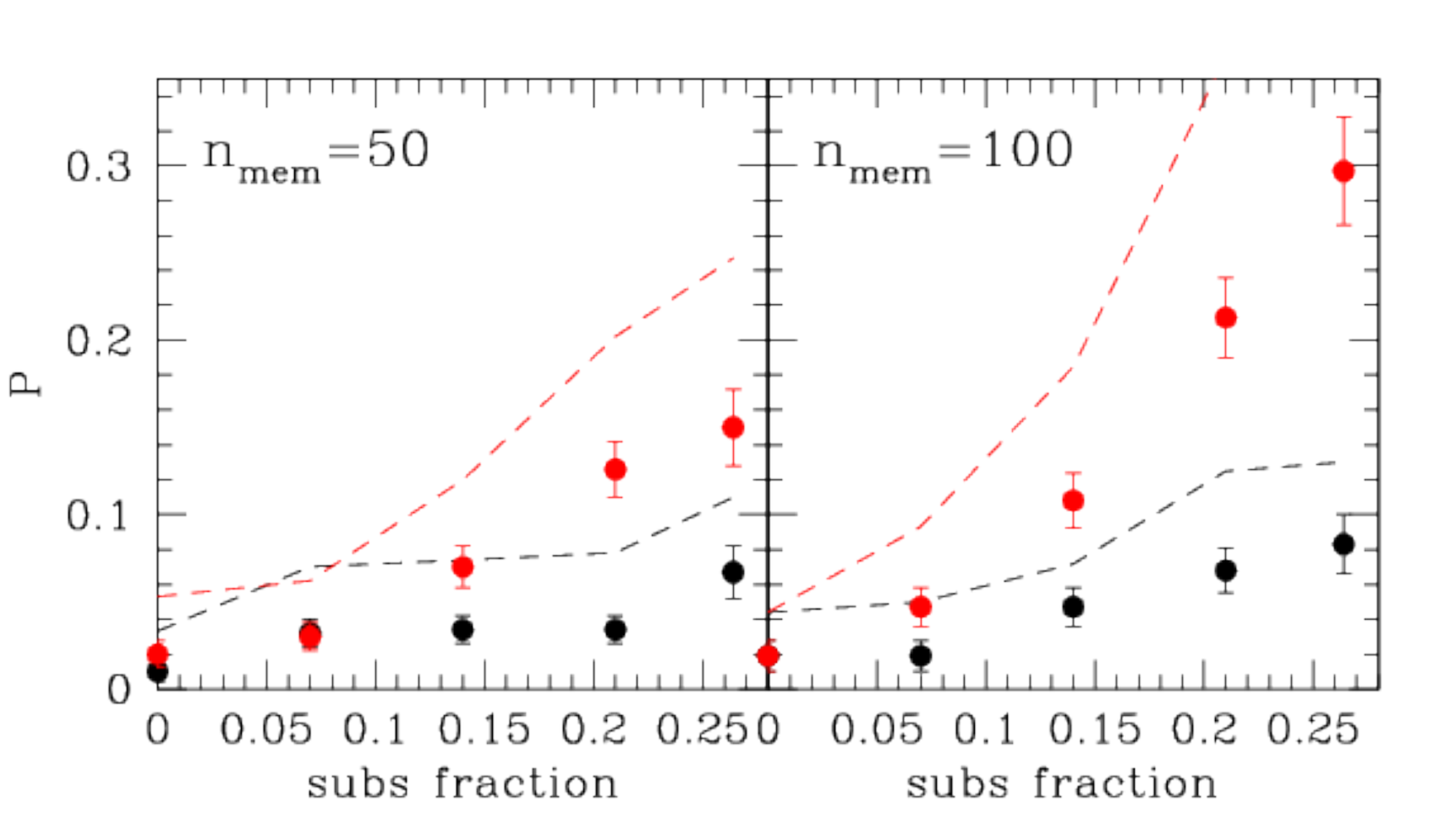}}
 \caption{The probability of misidentifying infalling substructures, if such exist,    for cluster rotation as a function of substructure richness (in    fraction of cluster members). With black we show results for the    case where the infall velocity is 50\% of the cluster velocity    dispersion, while with red we show the corresponding results for    the (more improbable) 100\% case. Dots show results based on the    strict criteria of rotation identification while dashed lines    show results based on the corresponding loose criteria.}
\end{figure}

\section{RESULTS}
\subsection{Individual cluster results}
\begin{figure*}
 \centering
 \resizebox{\hsize}{!}{\includegraphics[width=\textwidth]{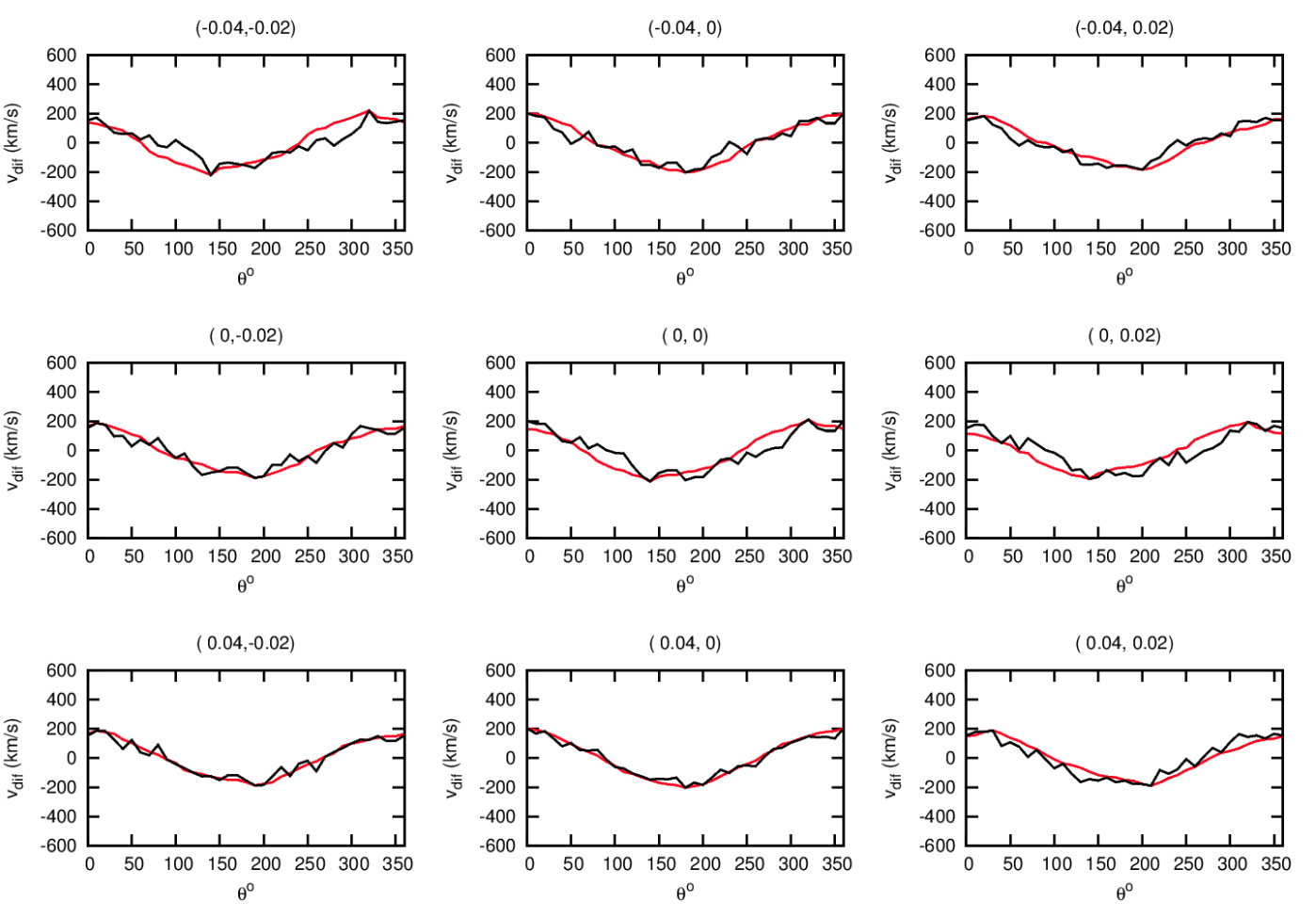}}
 \caption{The rotation diagrams for all the candidate rotational    centres for Abell 85 ($r<1.5 h^{-1}_{70}$ Mpc). Black lines are the    real rotation curves and red lines are the ideal rotation    curves. Above each panel we indicate the coordinates $(dy, dx)$ of    the rotational centre. The final selected one is that with $(dy,    dx)=(0.04, 0)$.}
 \label{A85cen}
\end{figure*}

Each cluster or subcluster in our sample is analysed in all four angular configurations according to the following sequence, the basic steps of which are already presented in section 4.2. First, we search for the best centre of possible rotation among nine tested (a related diagram for A85 is shown in Fig. \ref{A85cen} as an example). Using the selected centre, we apply our algorithm and construct the rotational diagram of the data,  the ideal rotation and that of the random velocity residuals, while we also construct the Kolmogorov-Smirnov (KS) probability curve as a function of rotation angle, $\theta$. The results of this analysis are then passed through the criteria discussed in Section 4.3.3 to decide whether a significant rotation has been detected, at any of the angular configurations of the cluster. 

In order to facilitate the visual verification of our results, we also construct for each cluster an aggregate plot with four panels, where we display:

(a) in the upper left panel the spatial distribution of the galaxies and their selected rotational centre, where residing and approaching galaxies are in red and blue colour, respectively, rejected galaxies due to velocity criteria are shown as black crosses, while rejected galaxies due to angular selection criteria as faint crosses,

(b) in the upper right panel the histogram of the line-of-sight galaxyvelocities along with the fitted Gaussian,

(c) in the lower left panel the data rotation diagram (points with errorbars), the ideal rotation (red continuous curve) and random rotation curves (blue continuous curve with dashed curves corresponding to 1$\sigma$ uncertainty), and

(d) in the bottom left panel the Kolmogorov-Smirnov probability diagram as a function of rotation angle $\theta$.

We will not present such diagrams for all the clusters of our sample, except for a few examples here and some interesting cases in the Appendix. As one example, we present for A85 the corresponding plots for the two main angular configurations ($r<1.5 h^{-1}_{70}$ and $r<2.5 h^{-1}_{70}$ Mpc). 

\begin{figure*}
 \resizebox{8.8cm}{8.8cm}{\includegraphics{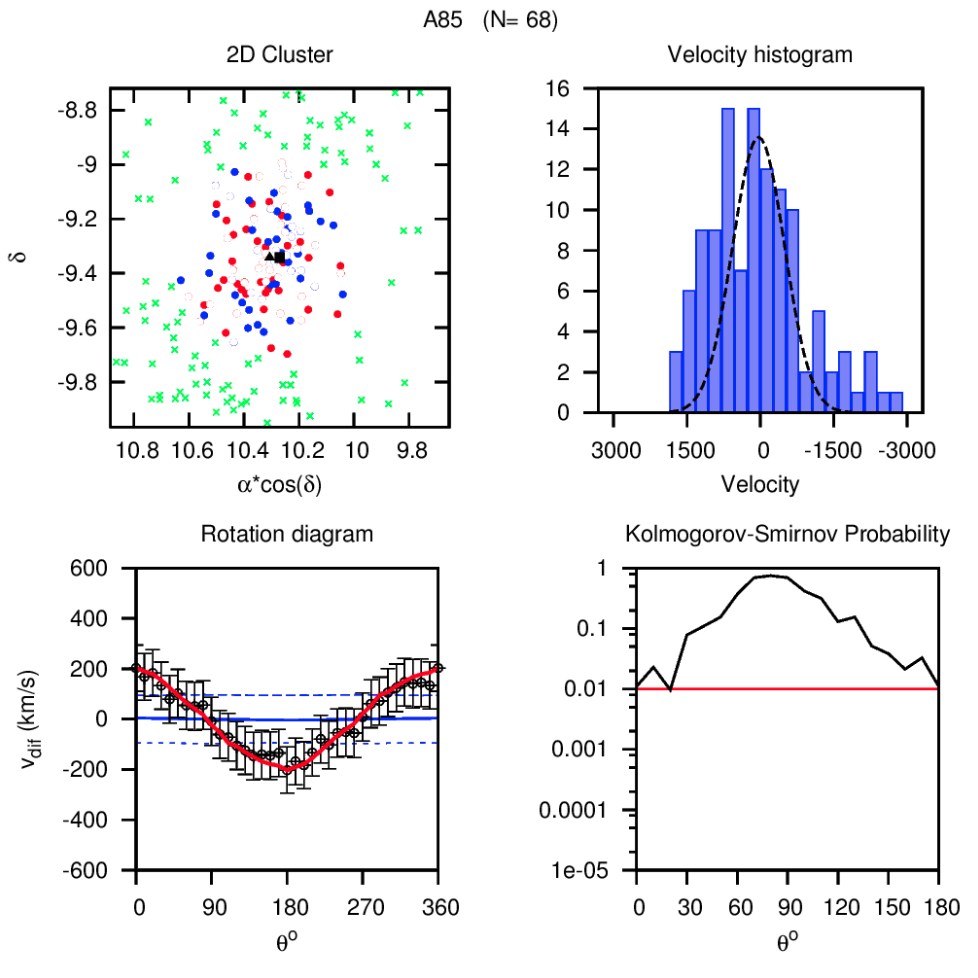}}
 \hfill
 \resizebox{8.8cm}{8.8cm}{\includegraphics{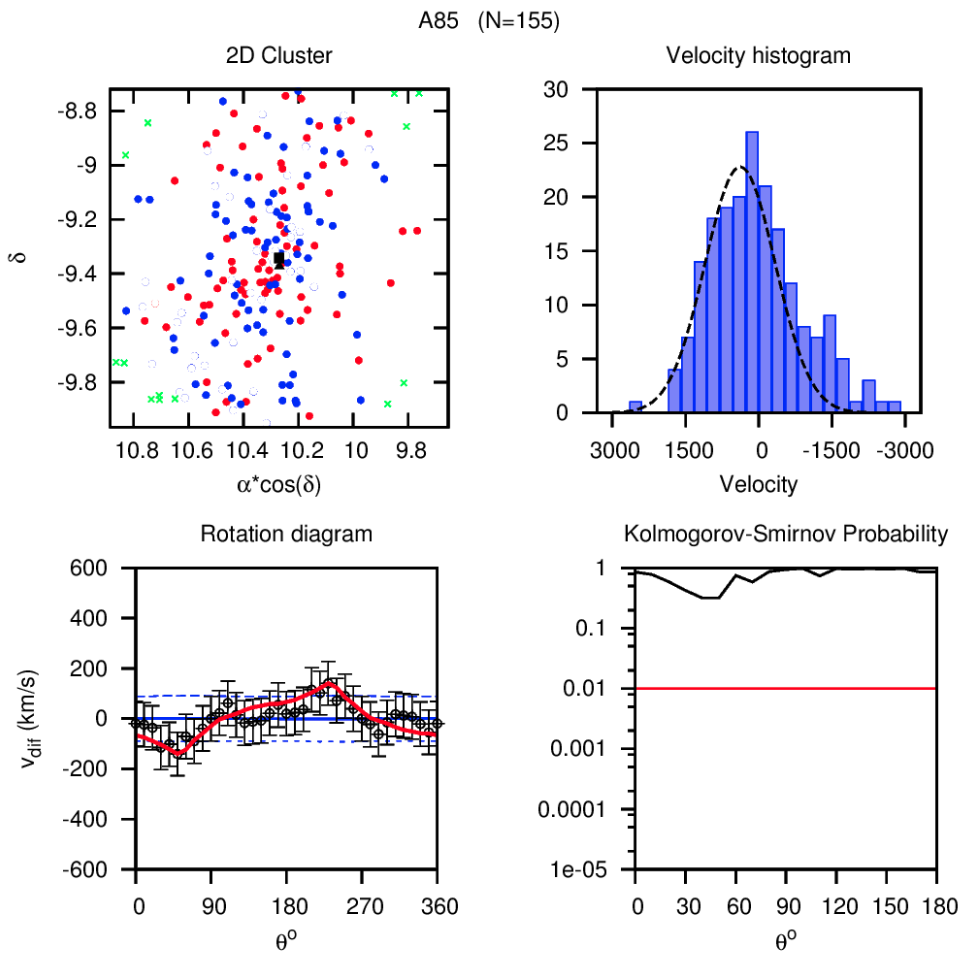}}
 \caption{The graphical outcome of the basic rotational diagram for    Abell 85. Within a radius of 1.5 $h^{-1}_{70}$ Mpc (left four panels)    and after excluding the outliers of the velocity distribution    (shown as empty points in the upper left panel),  the smooth sinusoidal data rotational diagram is evident, although it falls within the loose criteria. Within the 2.5 $h^{-1}_{70}$ Mpc (right four    panels) the rotational diagram and the $P_{KS}$ distribution are    consistent with no rotation.}
 \label{A85r}
\end{figure*}

We remind the reader that A85 is a rather rich BM type 1 cluster at a redshift $\langle z \rangle=0.055$, whose galaxy members with $m_r\lesssim 17.77$ (from SDSS DR10) vary between 68 and 155 at the two radii used. The relatively virialized nature of this cluster is confirmed by its smooth spherical X-ray profile \citep{ein}, although there are strong indications, when one goes to much fainter galaxies, of substructures \citep[][ and references therein]{a85}. However, if such substructures are manifested in the cluster velocity distribution they are already excluded by our ``cleaning'' procedure. Indeed there is such a case in A85, appearing  in velocity space at $|\langle v\rangle| \gtrsim \pm 1400$   km/sec (see the left velocity histogram in Fig. \ref{A85r}).

In Fig. \ref{A85r} we present the basic results of our analysis for the two cluster radii. Although for the $r=1.5 h^{-1}_{70}$ Mpc case there is a smooth sinusoidal rotation curve, exactly what expected for the ideal rotation (red curve), this cluster misses complying with the strict criteria of rotation, due to $\chi^2_{r}<1$. However, it complies with the loose criteria and thus it is considered as possibly rotating (but with a relatively low rotational velocity amplitude). When considering the larger cluster radius (right four panels of Fig. \ref{A85r}) we see that the indications of rotation vanish, a fact which could be due to small substructures acting as noise or due to a possibly different velocity distribution of the outskirt galaxies with respect to the inner ones; if for example they are infalling roughly isotropically to the cluster centre from the large-scale surrounding structure.

\begin{figure}
 \centering
 \resizebox{8.8cm}{8.8cm}{\includegraphics{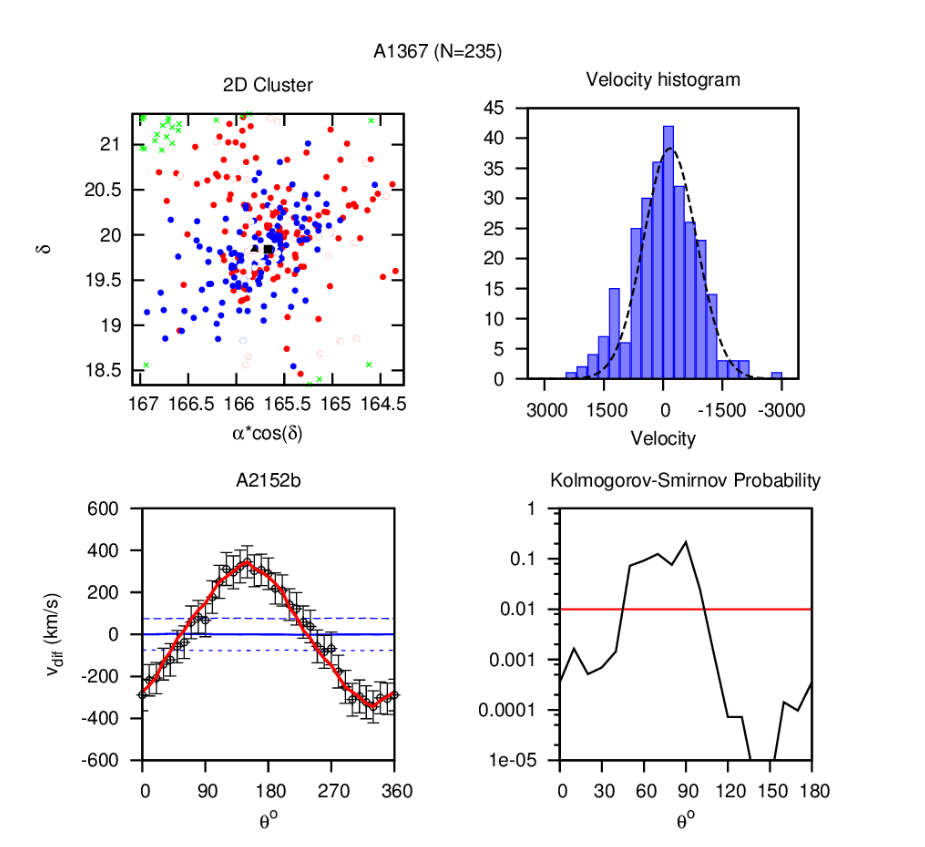}}
 \caption{The graphical outcome of the basic rotational diagram for    Abell 1367 within a radius of 2.5 $h^{-1}_{70}$ Mpc.    The excluded (velocity) outliers (corresponding to    known substructures) can be observed as empty circles in the upper left    panel. Based on the remaining galaxies, a clear and significant    sinusoidal rotational diagram is evident. }
 \label{A1367}
\end{figure}

To complete the presentation of some characteristic examples, we show in Fig. \ref{A1367} the relevant results for A1367, a rich and relatively nearby cluster ($z=0.022$). After excluding the outliers of the Gaussian fit to the galaxy velocity distribution (the galaxies at $<1300$ km/sec - see Fig. \ref{A1367}), we obtain what appears to be  a strongly rotating cluster showing a significant and unambiguous sinusoidal rotation diagram (in all four radial configurations). Although this cluster is known to show significant substructures in its central regions \citep{cortese}, we find even stronger rotational signals when excluding the central 0.3 or 0.5$h^{-1}$ Mpc region, an indication that although there are substructures, there is also rotation not necessarily attributed to coherent substructure velocity differences.

\subsection{Abell clusters with rotation}
Clusters for which we detect significant rotation, using either the strict or the loose criteria of rotation detection, and which have not been tagged as being dominated by substrutures (see section 4.1.2) are presented in Table \ref{t1} (for the $r=1.5 h^{-1}_{70}$ Mpc case), and table \ref{t2} (for the $r=2.5 h^{-1}_{70}$ Mpc case). In each table we also indicate clusters that show rotation only when excluding the inner cluster core (those with the star symbol), since projection effects are more severe along the central part of clusters, where typically a larger volume along the line of sight direction is sampled. The tables list the final number of galaxy cluster members selected, their mean redshift, the values of the four rotation indices discussed in section 4.3.3, the angle $\theta_{rot}$ of the rotation axis, which is the angle of the maximum semicircle mean velocity difference, the rotation amplitude, $v_{rot}$, which the maximum velocity difference in the rotation diagram, the cluster velocity dispersion, $\sigma_v$,  and the (crudely) corrected velocity dispersion after removing the cluster rotation (see section 5.3.4). 

\subsubsection{Effect of excluding the cluster core region}
Of the 14 clusters within $r=1.5 h^{-1}_{70}$ Mpc showing rotation under the strict criteria, two were detected only after excluding the inner $<0.3 h^{-1}$ Mpc core region (A2199, A2399), while only one cluster (A1913) is downgraded into the loose criteria rotation detection regime when excluding its inner core.

For the $r=2.5 h^{-1}_{70}$ Mpc case, out of the 19 clusters showing rotation under the strict criteria, 7 were detected only after excluding the inner $<0.5 h^{-1}_{70}$ Mpc core region. However, of these clusters four (A1913, A2089, A2147, and A2670) had originally been found rotating under the loose criteria. The only clusters found rotating under the strict criteria that lose completely their rotation when excluding the core region are: A426, A1228a and A1827. Other two clusters rotating under the strict criteria drop below the $n_{mem}=50$ limit, when excluding their core region, but retain their significant rotation detection (A1035a, A1291a). Similarly, out of the remaining five clusters with rotation under the loose criteria  (excluding the four that were upgraded to the strict regime when excluding the core region), one (A1238) was detected only when the core region was excluded. Finally, A1552 loses completely its  rotation when excluding the core region.

\begin{table*}
\begin{center} \caption{The clusters with significant rotation within   $r=1.5h^{-1}_{70}$ Mpc and with $n_{mem}\ge 50$, using either the   strict or loose criteria of rotation detection. The   first column is the Abell name of the cluster, the second is the   mean redshift of the members, the third is the number of members   used, the fourth is the orientation on the plane of the sky of   the rotation axis, the fifth is the rotation amplitude with its   uncertainty, the sixth and seventh are the coordinates of     the chosen rotation centre, the eighth is the minimum     value of the Kolmogorov-Smirnov probability, the next three   columns are $\chi^2_{id}$, $\chi^2_r$, $\chi^2_{id}/\chi^2_r$,   respectively, the twelfth is an indication for the  direction   of rotation (1 meaning clockwise and 2 anticlockwise). The last two   columns correspond to the initial and corrected, for rotation,   cluster velocity dispersion. Clusters that show significant rotation   only when excluding the inner cluster core  ($<0.3 h^{-1}_{70}$ Mpc) are   indicated with a star symbol.}
\label{t1}
\tabcolsep 3pt
\begin{tabular}[c]{lcrrcccccclccc}
Cluster & $z$ & $n_{mem}$ & $\theta_{rot}(^{\circ})$ & $v_{rot}/$km s$^{-1}$ & $\alpha_{cent}$ &$\delta_{cent}$ &$P_{KS}$ & $\chi^2_{id}/df$ & $\chi^2_r/df$ & $\chi^2_{id}/\chi^2_r$ & ${\cal I}$ & $\sigma_{v}($km s$^{-1})$ & $\sigma_{v, cor}($km s$^{-1})$ \\ \hline
 \multicolumn{14}{c}{Strict Criteria} \\ \hline

426	&	0.01729	&	136	&	10	&
498$\pm$128	&	37.178285	&	41.459958
&	0.000091	&	0.586	&	3.118	&	0.188
&	2	&	774	&	650	\\
1035a	&	0.06803	&	54	&	90	&
420$\pm$168	&	120.755407	&	40.185375
&	0.002970	&	0.168	&	1.862	&	0.09	&
2	&	566	&	461	\\
1169	&	0.05859	&	66	&	110	&	
473$\pm$134	&	120.234036	&	43.96349	&
0.000668	&	0.087	&	1.542	&	0.056	&
2	&	528	&	410	\\
1367	&	0.02124	&	177	&	130	&
271$\pm$90	&	165.754521	&	19.839167
&	0.005253	&	0.099	&	1.875	&	0.053
&	2	&	607	&	539	\\
1913	&	0.05303	&	102	&	40	&
348$\pm$108	&	207.638535	&	16.695574
&	0.001095	&	0.308	&	2.441	&	0.126
&	2	&	565	&	478	\\
2022 & 0.0581 & 51   & 90 & 362$\pm$107 &  198.830016  & 28.458548 &0.003460 & 0.117 & 1.603
& 0.073  & 2 & 403 & 311\\
2061	&	0.07878	&	74	&	230	&
218$\pm$88	&	198.151902	&	30.641212
&	0.003008	&	0.129	&	1.21	&	0.107
&	1	&	379	&	325	\\
2063	&	0.03457	&	102	&	140	&
441$\pm$149	&	228.196482	&	8.610653
&	0.005917	&	0.174	&	1.114	&	0.156
&	2	&	754	&	644	\\
2107	&	0.04127	&	111	&	160	&
403$\pm$117	&	218.143467	&	21.797129	&
0.001450	&	0.294	&	2.044	&	0.144	&	2
&	621	&	520	\\
2147	&	0.03573	&	223	&	130	&
303$\pm$102	&	231.317855	&	15.866537
&	0.000058	&	0.240	&	1.586	&	0.151
&	2	&	740	&	664	\\
2151	&	0.03665	&	175	&	250	&
514$\pm$93	&	229.879249	&	17.721073
&	0.000001	&	0.138	&	5.528	&	0.025
&	1	&	613	&	484	\\
2152	&	0.04442	&	85	&	140	&
286$\pm $76	&	231.466136	&	16.473324
&	0.000424	&	0.031	&	3.878	&	0.008
&	2	&	359	&	287	\\
2199*	&	0.03042	&	172	&	100	&
363$\pm$113	&	190.591969	&	39.524444
&	0.007317	&	0.178	&	1.934	&	0.092
&	2	&	730	&	639	\\
2399*	&	0.05743	&	68	&	280	&
333$\pm$108	&	326.378018	&	-7.776588
&	0.000330	&	0.287	&	1.74	&	0.165	&
1	&	443	&	360	\\ \hline

 \multicolumn{14}{c}{Loose Criteria} \\ \hline							
85	&	0.05518	&	68	&	0	&	214
$\pm$97	&	10.306381	&	-9.3425	&
0.009683	&	0.062	&	0.807	&	0.077	&
2	&	415	&	362	\\
1377	&	0.05194	&	62	&	110	&
429$\pm$166	&	99.476699	&	55.738889
&	0.000445	&	0.335	&	0.912	&	0.367
&	2	&	647	&	540	\\
2670	&	0.07608	&	91	&	250	&
394$\pm$170	&	352.6738	&	-10.405	&
0.005093	&	0.157	&	0.872	&	0.18	&
1	&	792	&	693	\\
1203*	&	0.07514	&	59	&	300	&
299$\pm$117	&	128.491013	&	40.294167
&	0.002565	&	0.048	&	0.91	&	0.053
&	1	&	442	&	368	\\
\end{tabular}

\end{center}
\end{table*}

\begin{table*}
\begin{center}
\caption{As in Table \ref{t1} but for clusters with significant   rotation within $r=2.5 h^{-1}_{70}$ Mpc. Clusters that show   significant rotation only when excluding the inner cluster core   ($<0.5 h^{-1}_{70}$) are indicated with a star symbol.}
\label{t2}
\tabcolsep 3pt
\begin{tabular}[c]{lcrrcccccclccc} 
Cluster & $z$ & $n_{mem}$ & $\theta_{rot}(^{\circ})$ & $v_{rot}/$km s$^{-1}$ & $\alpha_{cent}$ &$\delta_{cent}$ &$P_{KS}$ & $\chi^2_{id}/df$ & $\chi^2_r/df$ & $\chi^2_{id}/\chi^2_r$ & ${\cal I}$ & $\sigma_{v}($km s$^{-1}$) & $\sigma_{v, cor}($km s$^{-1}$) \\ \hline
 \multicolumn{14}{c}{Strict Criteria} \\ \hline
426	&	0.01722	&	155	&	20	&	405$\pm$122	&	37.315636	&	41.423283	&	0.007321	&	0.249	&	2.045	&	0.122	&	2	&	770	&	669	\\
1035a	&	0.06803	&	56	&	120	&	406$\pm$166	&	120.766955	&	40.185843	&	0.009703	&	0.283	&	1.885	&	0.15	&	2	&	559	&	458	\\
1228a	&	0.03521	&	65	&	70	&	157$\pm$53	&	140.673644	&	34.373472	&	0.006379	&	0.124	&	1.809	&	0.068	&	2	&	219	&	180	\\
1228b	&	0.04253	&	60	&	10	&	335$\pm$94	&	141.249465	&	34.190492	&	0.000019	&	0.411	&	3.077	&	0.133	&	2	&	322	&	239	\\
1291a	&	0.05087	&	50	&	30	&	382$\pm$103	&	96.411645	&	56.134474	&	0.000322	&	0.125	&	3.89	&	0.032	&	2	&	416	&	321	\\
1367	&	0.02148	&	237	&	150	&	354$\pm$75	&	165.810952	&	19.839167	&	0.000002	&	0.217	&	4.802	&	0.045	&	2	&	582	&	493	\\
1827	&	0.06516	&	50	&	300	&	190$\pm$93	&	194.65102	&	21.707222	&	0.001018	&	0.119	&	1.152	&	0.103	&	1	&	315	&	268	\\
2065	&	0.07224	&	170	&	70	&	712$\pm$176	&	204.198262	&	27.74665	&	0.000019	&	0.125	&	2.518	&	0.049	&	2	&	1166	&	988	\\
2151	&	0.03668	&	276	&	220	&	432$\pm$70	&	229.739375	&	17.748611	&	0	&	0.877	&	9.896	&	0.089	&	1	&	594	&	486	\\
2199	&	0.03057	&	344	&	80	&	325$\pm$77	&	190.728415	&	39.634998	&	0.000245	&	0.094	&	3.237	&	0.029	&	2	&	712	&	631	\\
2399	&	0.05754	&	103	&	250	&	201$\pm$85	&	326.401605	&	-7.764684	&	0.008962	&	0.177	&	1.134	&	0.156	&	1	&	428	&	378	\\
2152	&	0.04408	&	122	&	170	&	320$\pm$62	&	231.474846	&	16.403915	&	0.000095	&	0.379	&	6.029	&	0.063	&	2	&	374	&	294	\\
1185*	&	0.03362	&	140	&	330	&	292$\pm$89	&	147.127123	&	28.729761	&	0.001011	&	0.358	&	2.024	&	0.177	&	1	&	500	&	427	\\
1775a*	&	0.07523	&	57	&	160	&	308$\pm$112	&	184.073873	&	26.433694	&	0.00341	&	0.264	&	1.733	&	0.152	&	2	&	439	&	362	\\
1913*	&	0.05277	&	119	&	30	&	407$\pm$94	&	207.663393	&	16.708548	&	0.000142	&	0.745	&	4.214	&	0.177	&	2	&	536	&	435	\\
2022*	&	0.05798	&	53	&	50	&	423$\pm$103	&	198.830016	&	28.452586	&	0.000243	&	0.073	&	2.084	&	0.035	&	2	&	379	&	273	\\
2089*	&	0.07377	&	59	&	180	&	316$\pm$107	&	205.849001	&	28.039546	&	0.00044	&	0.064	&	1.136	&	0.056	&	1	&	431	&	352	\\
2147*	&	0.03624	&	327	&	230	&	304$\pm$95	&	231.281004	&	15.847379	&	0.000106	&	0.155	&	1.844	&	0.084	&	1	&	837	&	761	\\
2670*	&	0.07598	&	94	&	250	&
376$\pm$153	&	352.691984	&	-10.42811	& 0.008033	&	0.178	&	1.441	&	0.124	&
1	&	670	&	595	\\\hline
 \multicolumn{14}{c}{Loose Criteria} \\ \hline
1169	&	0.05887	&	83	&	120	&	248$\pm$147	&	120.30184	&	43.916405	&	0.009778	&	0.077	&	0.339	&	0.228	&	2	&	537	&	475	\\
1203	&	0.07527	&	89	&	300	&	244$\pm$92	&	128.476726	&	40.270752	&	0.00059	&	0.093	&	0.914	&	0.102	&	1	&	441	&	380	\\
1552	&	0.08611	&	104	&	200	&	261$\pm$125	&	183.577159	&	11.740556	&	0.002124	&	0.14	&	0.663	&	0.211	&	1	&	642	&	577	\\
1809	&	0.07911	&	88	&	170	&	272$\pm$102	&	207.530865	&	5.131838	&	0.002476	&	0.127	&	0.696	&	0.182	&	2	&	471	&	403	\\
1238*	&	0.07392	&	70	&	210	&	225$\pm$115	&	170.662776	&	1.068283	&	0.003468	&	0.173	&	0.82	&	0.212	&	1	&	487	&	431	\\

\end{tabular}
\end{center}
\end{table*}

\subsection{Statistical results}
In order to attempt to understand our results and possible causes of the cluster rotation, we will attempt to identify correlations between interesting cluster properties and rotation. To this end, we will use the Spearman correlation coefficient between any two parameters, $R_s$, and the probability that the correlations are consistent with the random expectation, ${\cal P}$. Positive $R_s$ means positive correlation, while negative $R_s$ means anticorrelation; a value near zero means the two parameters are not correlated. A small value of ${\cal P}$ indicates a significant correlation or anticorrelation at that level. We will report only relatively strong and relatively significant correlations, and as such we define: $|R_s|>0.3$ and ${\cal P}<0.05$.

\begin{figure}
 \centering
 \resizebox{\hsize}{!}{\includegraphics[width=.9\textwidth]{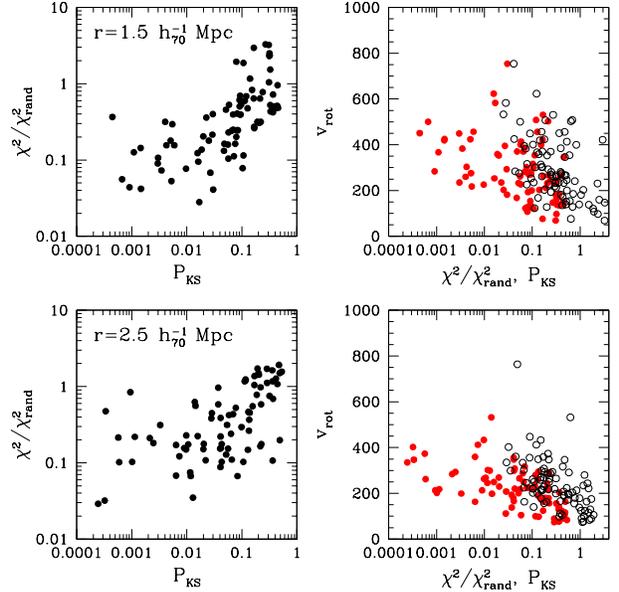}}
 \caption{Left-hand Panels: the scatter diagram between the two    rotation indices (upper for the $r=1.5 h^{-1}_{70}$ Mpc case and    lower for the $r=2.5 h^{-1}_{70}$ Mpc case). Right-hand Panels:    tThe rotation amplitude, $v_{rot}$, as a function of the  Kolmogorov-Smirnov probability (red filled symbols) and as a    function of the $\chi^2_{min}$ ratio value (empty black symbols).} 
\label{25ind}
\end{figure}

A first observation is that all the indices that we use to deduce rotation are correlated strongly among them,  as it can be seen in Fig. \ref{25ind}, where we plot only clusters that have not been excluded from the analysis due to strong substructuring (see section 4.1.2). The value of the Kolmogorov-Smirnov probability, $P_{KS}$, and the value of the ratio of the $\chi^2$ minimium values between the ideal and real rotation diagrams are strongly correlated with each other in all angular configurations. For example, for the $r=1.5 h^{-1}$ Mpc case we obtain $R_s=0.75$ and ${\cal P}<10^{-10}$. Also the amplitude of the rotation, $v_{rot}$, is strongly correlated with both rotation  indices with $R_s\gtrsim 0.6$ and ${\cal P}<10^{-9}$.  This also should be expected since when the rotational velocity is large, the rotation will be more clearly identified, and vice-versa. 

\subsubsection{Check for systematic biases of the rotation indices}
Before we present our main results it is important to make sure that we understand the possible systematic effects of the resulting rotation indices for the clusters studied. We have already investigated and quantified the effect of shot-noise and undetected substructures (section 4.3.4), however, we further check for correlation of the resulting rotation indices on the number of galaxy members, $n_{mem}$, and on $z$. As we already showed, $n_{mem}$ needs to be relatively large in order to unambiguously detect a rotation if such exist. However, since the input galaxy catalogue (the SDSS DR10 spectroscopic catalogue) is limited to $m_r\sim 17.77$, when we look at larger distances we observe less and brighter galaxy cluster members. Therefore, there will be an unavoidable redshift dependence of the cluster galaxy membership and thus a redshift dependence of the rotation indices is possible. This does not necessarily imply an important problem but rather that the fraction of rotating clusters found should be considered a lower limit.

In any case, we have tested for such a dependence for the $n_{mem}\ge 50$ case and we find weak and marginally significant  correlations, in any case below the limit we set in section 5.2. Only in the case of the Kolmogorov-Smirnov test,  which due to its nature a dependence of $P_{KS}$ on  $n_{mem}$ is expected and, since the latter is anticorrelated with $z$ (as discussed previously), we expect a correlation of $P_{KS}$ with $z$. Indeed we find such a weak but relatively significant dependence ($R_s=0.33$ and ${\cal P}\simeq 0.013$)  for the $r=1.5 h^{-1}_{70}$ Mpc and $n_{mem}\ge 50$ case. For the $r=2.5 h^{-1}_{70}$ Mpc case the above correlation becomes weaker.

\subsubsection{Fraction of rotating clusters} In this section we present some basic statistics regarding the fraction of clusters that show indications of rotation, based on both strict and loose rotation criteria, as defined in the previous section. In Table \ref{fract} we present the number of clusters and the corresponding fraction of the total that show   strict or loose indications of rotation for clusters with $n_{mem}\ge 50$, for which the rotation identification is quite secure. The fractions are slightly different when limiting the studied area within 1.5 or 2.5 $h^{-1}_{70}$ Mpc of the cluster centre, with the latter being slightly smaller than the former. We  also present the final overall number of unique clusters rotating using any of the four spatial configurations, as discussed below.
 
Overall, it is secure to say that Abell clusters with $n_{mem}\ge 50$ showing significant indications of rotation, within either of the two limiting radii, range between $\sim 25\%$ (for the strict criteria)  and $\sim 32\%$ (for the loose criteria) of the total.

It should be noted however, that the specific clusters showing rotation at the different radii are not always the same. In particular, out of the 18 rotating clusters of Table \ref{t1}, five are missed when using $r=2.5 h^{-1}_{70}$ Mpc. Also, quite a few more clusters appear to be rotating when we extend our analysis to $r=2.5 h^{-1}_{70}$ Mpc than within $r=1.5 h^{-1}_{70}$ Mpc. In particular, out of the 24 clusters listed in Table \ref{t2} only 13 are  found rotating within $r=1.5 h^{-1}_{70}$ Mpc, taking also into account one that drops below the $n_{mem}=50$ limit.
\begin{table}
\begin{center}
\caption{Fraction of clusters showing rotation under     the strict and loose criteria, for the analysed     clusters with $\ge 50$ members (which are less prone to   random errors). The final, corrected for the expected number of   false detections according to our simulations, fraction of rotating   clusters is also listed.}
\label{fract}
\tabcolsep 6pt
\begin{tabular}{cccc} 
Radius/$h^{-1}_{70}$Mpc & $N_{clus}$ & Strict &
Loose \\ \hline
1.5 & 56  & 14 ($25\%$) & 18 ($32\%$) \\
2.5 & 86  & 19 ($22\%$) & 24 ($28\%$) \\ \\
Overall & 86 & 23 ($27\%$) & 29 ($34\%$)\\ 
Corrected & 86 & 23\% & 28\% \\ \hline
\end{tabular}
\end{center}
\end{table}

We can reach an overall number, and the corresponding final fraction of rotating clusters with $n_{mem}\ge 50$ at any of the two radii and taking also into account results based on excluding the core region. Using the strict criteria we find 23 such clusters, corresponding to $\sim 27\%$ of the total (86), while using the   loose criteria we find 29 such clusters, corresponding to  $\sim 34\%$ of total.

\subsubsection{Fraction of clockwise and anticlockwise rotating   clusters}
As we discussed in section 2.1 our algorithm provides us with the direction of rotation for the rotating clusters. What is expected in an initially irrotational Universe on large-scales is the lack of a preferred direction of cluster rotation. In Tables 2 and 3 we present the direction of rotation for each of our rotating clusters, indicated by the symbol ${\cal I}$ which takes the value 1 for rotation or 2 for counter rotation.  Using only the results based on the strict criteria,  we obtain for the $r=1.5 h^{-1}_{70}$ Mpc case  11 anticlockwise and only 3 clockwise rotating clusters, while for the $r=2.5 h^{-1}_{70}$ Mpc case we have 12 and 7, respectively. There appears to be a slight preference for clockwise rotating clusters\footnote{See \citet{longo} for a similar results on spiral galaxy rotation.}, but what is the significance  of the number difference, $\Delta=8$ for the former case and $\Delta=5$ for the latter case? The Poisson uncertainty of $\Delta$ is $\sigma_{\Delta}\simeq 3.7$ and 4.4 for the two cases, respectively, a fact which implies that the difference is significant only at a 2.1$\sigma$ and 1.1$\sigma$ level, respectively. We do not consider as overwhelming the former significance and we conclude that there is no significant evidence for a preferred  direction of rotation among the rotating clusters.

\subsubsection{Correcting the cluster velocity dispersion for rotation}
In order to correct the cluster velocity dispersion, assumed to be due to roughly isotropic galaxy orbits, for the rotational modes, we assume that the two velocity components are independent and that the  expected cluster velocity dispersion due to the rotational velocity $v_{rot}$ can be approximated, assuming an ideal rotation, as:
 \begin{equation}
\sigma^2_{rot}\simeq 2 \times \left(\frac{v_{rot}}{2}\right)^2\;.
\end{equation}
Therefore, the corrected cluster velocity dispersion, $\sigma_{v,cor}$, is approximately provided by the following:
\begin{equation}
\sigma^2_{v,cor}\simeq
\sigma_{v,raw}^2-\sigma^2_{rot}=\sigma_{v,raw}^2-\frac{v_{rot}^2}{2}\;.
\end{equation}

For the majority of the rotating clusters, the corrected velocity dispersion is not dramatically altered,  but the correction is not insignificant. Defining the fractional difference between corrected and uncorrected cluster velocity dispersion as $$\delta \sigma_v=\frac{\sigma_{v,raw}-\sigma_{v,cor}}{\sigma_{v,raw}}$$ we obtain for the $r=1.5 h^{-1}_{70}$ Mpc case a median value of $\sim$10\%, and a mean value of $\langle \delta \sigma_v\rangle \sim 12\%$. A similar analysis for cluster rotation out to $r=2.5 h^{-1}_{70}$ Mpc provides the following fractional differences: a median value of $\sim$12\%, and a mean value of $\langle \delta \sigma_v\rangle \sim 15\%$. The corresponding corrected cluster mass is given by: $$ M_{cor}\simeq M_{raw} (1-\delta\sigma_v)^2 \;,$$ implying a corrected cluster mass reduced by 20\% - 30\% on average, with respect to that uncorrected for rotation.

\subsubsection{Correlations between cluster rotation and cluster physical parameters}
We attempt to investigate whether there are correlations between the rotation indices and the different physical properties of the clusters, dynamical or other. We correlate the two main rotation indices, ie., $\chi^2_{id}/\chi^2_r$ and $P_{KS}$ with the characteristics of the cluster dynamical state, ie., BM type and $X_r$ (defined in section 4.1.4), and with the cluster mass, characterized either by the number of bright galaxies, $N_{*}$, or by the cluster velocity dispersion, $\sigma_v$ (defined in section 4.1.3).
 
For both radii we find no significant correlations between cluster mass or cluster dynamical state and rotation indices. However, since the majority of clusters do not show signs of rotation they would act as noise weakening possible correlations between rotation and cluster parameters for those clusters that show significant indications of rotation.  Indeed, selecting only the latter clusters we find relatively significant correlations but only between the strength and significance of rotation and the cluster dynamical state (not with cluster mass), in the direction of a correlation between rotation strength and dynamical youth (see table \ref{Spear}).When we analyse clusters that show rotation for the $r=2.5 h^{-1}_{70}$ Mpc case, we find correlations  only between BM or $X_s$ and $P_{KS}$, ie., not with the prime indicator of rotation ($\chi^2_{id}/chi^2_{ran}$), but only with the significance of the semicircle velocity difference.

We can conclude that there are indications that the cluster rotation is related to the earlier phases of cluster virialization but after the initial anisotropic accretion and merging has taken place (since we have excluded all clusters showing significant substructure in projected or velocity space).

\begin{table}
\begin{center}
\caption{Spearman's correlation coefficient and the probability that   the detected correlation is consistent with the random expectation   for the indicated pairs of parameters using rotating, under  trict rotation criteria, clusters with $n_{mem}\ge 30$.}
\label{Spear}
\tabcolsep 3pt
\begin{tabular}{ccccccc} 
 & \multicolumn{3}{c}{$r=1.5 h^{-1}$ Mpc} & \multicolumn{3}{c}{$r=2.5
    h^{-1}$ Mpc} \\ \hline
       & BM-$P_{ks}$ & BM-$\chi^2_{id}/\chi^2_r$  & $X_s-\chi^2_{id}/\chi^2_r$ & &BM-$P_{ks}$ & $X_s-P_{ks}$ \\ \hline
$N$    & 15         &    15        &    12  & &  20   &  15  \\
$R_s$  & -0.49      & -0.44        & -0.62  & &-0.41  & -0.52 \\
${\cal P}$  & 0.062  & 0.096       & 0.033  & &0.076  & 0.045 \\
\end{tabular}
\end{center}
\end{table}

\section{Conclusions}
We searched for possible cluster rotation in a sample of Abell clusters using galaxy-member redshifts from the SDSS DR10 spectroscopic data base. We developed a new algorithm in order to be able to deduce rotation using the line-of-sight velocities of the galaxy members. We verified the performance of this algorithm by applying it on various Monte Carlo simulated clusters with known rotational characteristics. We also compared our method with that of the \citet{hwanglee} method. 

Our algorithm provides the significance of the rotation identification (with a set of indices), the rotation amplitude, the position angle of the rotation axis, whether the rotation is clock or anticlockwise and the rotation centre. We find that the amplitude of the rotation is correlated with the indications of rotation; the larger the rotation amplitude the more significant are the indications of rotation. This implies that small amplitude rotation may not be easy to identify, and thus it could pass undetected.

We then applied our algorithm on our sample of Abell clusters using two different sets of criteria for rotation identification, the so-called strict and loose criteria and two different outer cluster radii (1.5 and 2.5 $h^{-1}_{70}$ Mpc). Out of 86 cluster with more than 50 member galaxies we have found in total 23 rotating clusters (in any of the 2 radii studied) using the strict criteria of rotation identification and 29 such clusters using the loose criteria of rotation identification. Taking into account the expected fraction ($\sim 10\% - 15\%$) of misidentified coherent substructure velocities for rotation, provided by our Monte Carlo simulation analysis, the corresponding final fraction of rotating clusters is $\sim 23\%$ and $\sim 28\%$, respectively, under  the strict and loose criteria. These results appear to be in tension with recent numerical $N$-body simulations \citep{baldi} which find a significantly smaller fraction of rotating clusters; however with slightly different criteria of rotation.

Finally, when we use the inner radius case (1.5 $h^{-1}_{70}$ Mpc) and clusters that show indications for rotation, we find relatively significant correlations between the cluster dynamical state (X-ray isophotal shape as well as the BM type) and the significance  of cluster rotation, a fact which implies that the cluster rotation could be related to the dynamically younger phases of cluster formation but after the initial anisotropic accretion and merging have taken place. This hints towards the inner radius rotation being related to the initial anisotropically accreted matter having significant angular momentum, which gets amplified by collapse. The fact that we find  fewer such correlations when we use clusters with rotation within the outer cluster radius (2.5 $h^{-1}_{70}$ Mpc) possibly hints towards a different cause or a different phase of the relevant rotation, possibly being related to the imprint of coherent rotational motions of galaxies in the cluster outskirts prior to dynamically disturbing the cluster inner regions.

\section*{Acknowledgements}
MP would like to thank Hrant Tovmassian for suggesting the study of cluster rotation and for many initial discussions on the subject.

\bibliographystyle{mn2e}
\bibliography{ms8}

\appendix
\section{Clusters successfully divided in substrustures}
We list here those clusters of our sample that were found to consist in velocity space of two or more well-separated substructures. These clusters were separated into their different components, which were individually analysed for rotation when possible.

\begin{itemize}

\item \textbf{Abell 1035} \\
This cluster presents a background subcluster in all four configurations studied. One of the two subclusters was found to have a significant rotational mode.\\

\item \textbf{Abell 1228} \\
Abell 1228 was found to consist of three well-separated components in velocity space aligned along the line-of-sight, in all four spatial configurations (see Fig. 8). Two components are rich enough to be analysed for rotation and indeed they show strong indications of rotation, in the 2.5 $h^{-1}_{70}$ Mpc and 0.5-2.5 $h^{-1}_{70}$Mpc configurations, with rotational velocity amplitude of   $\sim 200$ km/s (A1228a) and $\sim 400$ km/s (A1228b). The two  subclusters rotate in the same direction (${\cal I}=2$) but have their (projected on the plane of the sky) rotation axes perpendicular to each other (figure A1). \\

\item \textbf{Abell 1291} \\
Another interesting case is Abell 1291. Studying its galaxy member velocity distribution we again identify 3 different peaks, clearly separated from each other. The third and most distant substructure could not be studied due to its small richness. From the other two only the nearest one (A1291a) show indications of rotation  for the 2.5 $h^{-1}_{70}$ Mpc and 0.5-2.5 $h^{-1}_{70}$ Mpc configurations. \\

\item \textbf{Abell 1775} \\
We found a foreground group of galaxies in velocity space and in all the four spatial configurations. This substructure is placed south-east of the main cluster of galaxies and was not found to present any indications of rotation in any of the configurations, while the main cluster is found to rotate in the 0.5-2.5 $h^{-1}_{70}$ Mpc configuration.\\

\item \textbf{Abell 2152} \\
Abell 2152 presents a main group of galaxies found to have strong indications of rotation in all configurations. In all configurations foreground galaxies (part of the Hercules supercluster) are found with a wide velocity distribution, which we succesfully exclude from our analysis. The main cluster is then found to have a significant rotational mode. \\

\begin{figure}
 \centering
 \resizebox{\hsize}{!}{\includegraphics[width=.45\textwidth]{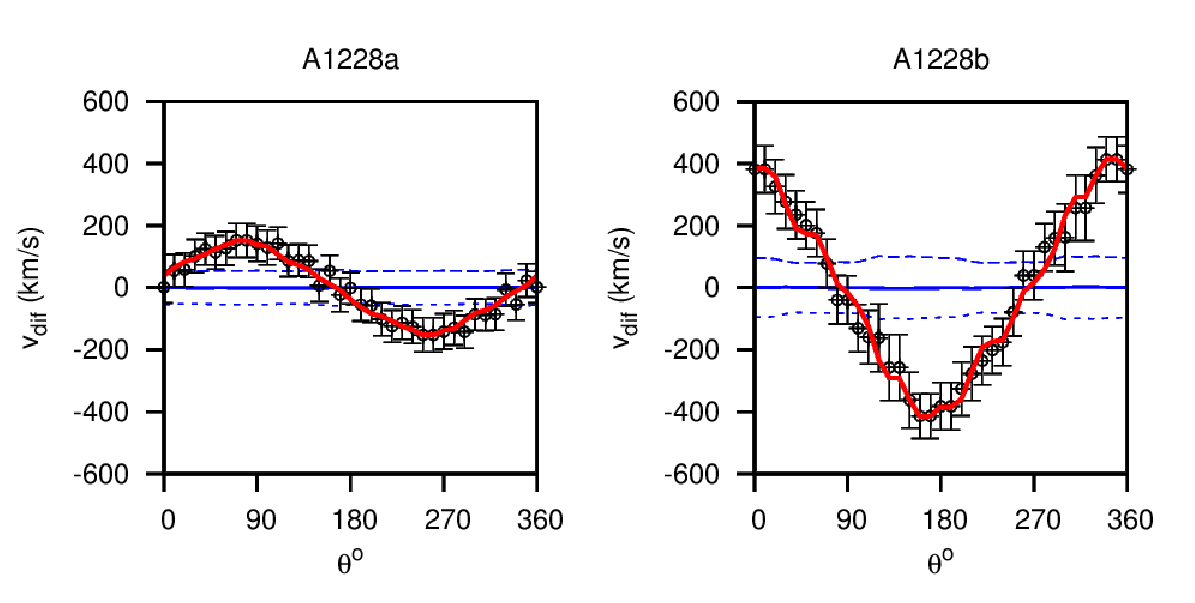}}
 \caption{The rotational diagram for    Abell 1228a (left) and 1228b (right) within a radius of 2.5    $h^{-1}_{70}$ Mpc. The two subclusters have perpendicular rotation    axes.}
 \label{A1228ab}
\end{figure} 

\end{itemize} 

\end{document}